\documentstyle[prb,aps,eqsecnum,epsfig,multicol]{revtex}
\def\ltsim{\mathop{\raise3pt\hbox{$<$}\llap{\lower3pt\hbox{$\sim
$}}}}
\def\gtsim{\mathop{\raise3pt\hbox{$>$}\llap{\lower3pt\hbox{$\sim
$}}}}
\newcommand{\be}{\begin{eqnarray}}
\newcommand{\ee}{\end{eqnarray}}

\newcommand{\bra}[1]{
\left\langle \, #1                \, \right|
}
\newcommand{\ket}[1]{
\left|       \, #1                \, \right\rangle
}

\begin{document}
\draft

\title{Single hole dynamics in the $t$-$J$ model on a square lattice}
\author{Michael Brunner, Fakher F. Assaad, and Alejandro Muramatsu}
\address{
Institut f\"ur Theoretische Physik III, Universit\"at Stuttgart,
Pfaffenwaldring 57, D-70550 Stuttgart,\\ 
Federal Republic of Germany
}
\date{\today}
\maketitle
\begin{abstract}
We present quantum Monte Carlo (QMC) simulations for a single hole in a 
$t$-$J$ model from $J=0.4t$ to $J=4t$ on square
lattices with up to $24 \times 24$ sites. 
The lower edge of the spectrum is directly extracted
from the imaginary time 
Green's function.
In agreement with earlier 
calculations, we find flat bands around $(0,\pm\pi)$, $(\pm\pi,0)$
and the minimum of the dispersion at $(\pm\pi/2,\pm\pi/2)$.
For small $J$ both self-consistent Born approximation and series
expansions give a bandwidth for the lower edge of the spectrum
in agreement with the simulations, whereas for $J/t > 1$, only
series expansions agree quantitatively with our QMC results. 
This band corresponds to a coherent 
quasiparticle.
This is shown by a finite size scaling of the quasiparticle 
weight $Z(\vec k)$ that leads to a finite result in the 
thermodynamic limit for the considered values of $J/t$.
The spectral function
$A(\vec k, \omega)$ is obtained from the imaginary time Green's
function via the maximum entropy method. 
Resonances above the
lowest edge of the spectrum are identified, whose 
$J$-dependence is quantitatively described by string excitations
up to $J/t=2$. 
\end{abstract}
%
\begin{multicols}{2}
\section{Introduction}
Since the pioneering work by Brinkman and Rice \cite{brinkman70} (BR)
the dynamics of a hole in an antiferromagnet remained as a recurring 
open problem in condensed matter physics. After the discovery of 
high temperature superconductors \cite{bednorz86} and the suggestions 
by Anderson \cite{anderson87}
on the possibility of a non-Fermi liquid state in those materials,
the question whether the quasiparticle weight
of a hole vanishes due to the 
interaction with an antiferromagnetic background became central in the 
field of strongly correlated fermions. 
 
The BR treatment led to a fully incoherent spectrum in the so-called
retraceable path approximation, for an antiferromagnetic Ising-like 
background, in the limit
$J_z \rightarrow 0$. The retraceable path approximation is exact in one 
\cite{brinkman70} and in infinite dimensions \cite{Metzner92} but
not in two dimensions since contributions of loops 
(Trugman paths \cite{trugman88}) may lead to a coherent propagation 
of the hole. Furthermore, for an Ising-like background it was shown 
within a Lanczos scheme \cite{Zhong95}, that a finite quasiparticle
weight is obtained.
For the case of physical interest, namely with
a Heisenberg spin-background, a large number of
numerical methods \cite{dagotto94} led to conflicting results. Whereas 
exact diagonalizations
found large quasiparticle peaks at the lower 
edge of the spectrum \cite{poilblanc93}, quantum Monte Carlo (QMC)
results were interpreted as leading to a vanishing quasiparticle 
weight \cite{Sorella92b}. 
Since exact diagonalizations are possible only
on very small lattices, finite
size scaling cannot be performed reliably. On the other hand, QMC
simulations suffered from the minus-sign problem, such that
scaling was not possible with reasonable confidence.
Further studies based on the self-consistent
Born approximation (SCBA) \cite{kane89,martinez91,liu92}
gave a finite quasiparticle weight. However, since there 
fluctuations of the spin-background are only taken into account
in the frame of a spin-wave approximation, the results obtained are not 
conclusive. Exact results for the supersymmetric point $J=2t$ were
obtained by Sorella \cite{sorella96b}, that
give important benchmarks for any analytical
or numerical method (see Sec.~\ref{sec:qp}), 
but unfortunately, they cannot be rigorously extended
to the physical relevant parameter range $J \sim 0.4t$.

Quite recently, the dynamics of a single hole in an antiferromagnetic
background became experimentally accessible by
angle resolved photoemission spectroscopy 
(ARPES) in undoped materials like 
Sr$_2$CuO$_2$Cl$_2$ \cite{wells95,Kim98} 
and Ca$_2$CuO$_2$Cl$_2$
\cite{Ronning98}. The main features observed there are a minimum of the 
dispersion at $\vec k =(\pi/2,\pi/2)$ together with a vanishing of
spectral 
weight beyond this point along the (1,1) direction. The obtained spectra 
show that the very flat portion around $(\pi,0)$, that in optimally doped 
materials is almost degenerate with the bottom of the spectrum at 
$(\pi/2,\pi/2)$ \cite{marshall96}, is shifted upwards (in a hole 
representation) by approximately 300 meV. This contradicts 
the single hole spectra found theoretically so far, where essentially
the lower edge of the spectrum at $\vec k =(\pi/2,\pi/2)$ and 
$(\pi,0)$ are almost degenerate, such that additional second and third
nearest neighbor hopping terms were suggested \cite{Kim98,Tohyama99}, that
lead to an agreement of the exact diagonalization
results with experiments. 
Such terms
were made recently responsible also for the vanishing of spectral weight
close to $(\pi/2,\pi/2)$ by reducing the quasiparticle 
weight \cite{Tohyama99,Martins99}.  

In the following we present dynamical properties of a single hole in 
a two-dimensional $t$-$J$ model on lattices with up to $24 \times 24$
sites in the parameter range $0.4 \leq J/t \leq 4$.
Results  were obtained 
with a new QMC algorithm, where the spin background is simulated with
a loop-algorithm \cite{evertz93} and the hole is exactly propagated 
for a given configuration of the spin background. The lower edge of
the spectrum is obtained directly from the asymptotic form of the
imaginary time Green's function. The resulting dispersion agrees with
previous results obtained within SCBA and
series expansions \cite{hamer98} 
for $J/t < 1$, whereas for $J/t > 1$ only agreement with 
series expansions is found.
In particular, a flat dispersion is obtained around $\vec k = (\pi,0)$
very close in value to the bottom of the band at $\vec k = (\pi/2,\pi/2)$,
in contrast to the experiments \cite{wells95,Kim98,Ronning98}.
The asymptotics of the imaginary time Green's function delivers also the 
quasiparticle weight for that band. Finite size 
scaling is presented showing that $Z(\vec k)$ is finite for the parameter 
range considered, such that the lower edge of the spectrum corresponds 
to a coherent quasiparticle.
Furthermore, our data are consistent with
another exact prediction \cite{sorella96b}, 
namely that at the supersymmetric point and in the thermodynamic limit, 
$Z({\vec Q})/Z(0) = (2m)^2$, where $\vec Q = (\pi,\pi)$ is the 
antiferromagnetic wave vector and $m$ is the staggered
magnetization.
The spectral function $A (\vec k, \omega)$ is calculated by analytic 
continuation with maximum entropy (MaxEnt) \cite{jarrell96}. 
Overall agreement is 
found with exact diagonalizations. At the supersymmetric point, the 
delta function predicted by Sorella \cite{sorella96b} for the wave vector 
$\vec k = (0,0)$ is {exactly} reproduced. 
By extracting the contribution of the quasiparticle from the imaginary
time Green's function, a resonance above the quasiparticle band is made 
evident, that together with the lower edge of the spectrum scales as 
$J^{2/3}$, in agreement with the string picture
\cite{bulaevskii68} used to described the 
excitations for a hole in an antiferromagnetic Ising-background. 
Remarkably, also
the prefactors of the corresponding Airy functions are needed in order to 
properly describe the distance between the resonance and
the quasiparticle band.

The paper is organized as follows.
Section\ \ref{sec:alg} describes the model, a 
canonical transformation that leads to a bilinear form in 
spinless fermions
interacting with $S=\frac{1}{2}$ pseudospins,
and the algorithm. Since the 
Hamiltonian for the transformed $t$-$J$ model is bilinear in fermions
(the holes), their propagation can be calculated exactly given a 
pseudospin configuration. In Sec.\ \ref{sec:res}
the results are discussed. 
Section\ \ref{sec:low} describes the lower edge
of the spectrum and how it is
obtained. In Sec.\ \ref{sec:qp}
the results for the quasiparticle weight are
shown. Section\ \ref{sec:sf} describes  the spectral 
function $A(\vec k, \omega)$ and the string excitations. Finally, 
the conclusions are given in Sec.\ \ref{sec:conc}. 
\section{The model and the algorithm}
\label{sec:alg}
The $t$-$J$ model is a suitable one to simulate the dynamics of a 
single hole in an antiferromagnet. On the one side, it can be obtained 
from the Hubbard model in the large coupling limit, which at half-filling 
leads to the Heisenberg antiferromagnet. On the other side, it is the 
relevant one to simulate the cuprates, as shown by Zhang and Rice 
\cite{zhang88}, and hence, to compare with experiments 
\cite{wells95,Kim98,Ronning98}.
Its Hamiltonian is
\begin{equation}
H_{t-J}=
-t \sum\limits_{<i,j>,\sigma} \tilde c^{\dagger}_{i,\sigma} \tilde
c^{}_{j,\sigma}
+J \sum\limits_{<i,j>}
\left( \vec S_i\cdot \vec S_j -\frac 1 4 \tilde n_i \tilde n_j \right)\; ,
\label{glg:tJ}
\end{equation}
where $\tilde c^{\dagger}_{i,\sigma}$ are projected fermion operators
$\tilde c^{\dagger}_{i,\sigma}=(1-c^{\dagger}_{i,-\sigma}
c^{}_{i,-\sigma})c^{\dagger}_{i,\sigma}$
,
$\tilde n_i=\sum\limits_{\alpha} \tilde c^{\dagger}_{i,\alpha}\tilde
c^{}_{i,\alpha}$,
$\vec S_i=(1/2)\sum\limits_{\alpha,\beta}c^{\dagger}_{i,\alpha}
\vec{\sigma}_{\alpha,\beta}c^{}_{i,\beta}$,
and the sum runs over nearest neighbours only.
In order to render Eq.~(\ref{glg:tJ}) a bilinear
form in fermionic operators,
we perform  a canonical transformation \cite{khaliullin90}
\begin{equation}
c^\dagger_{i\uparrow} = \gamma_{i,+} f_i - \gamma_{i,-} f_i^\dagger \, ,
\; \; \;
c^\dagger_{i\downarrow} = \sigma_{i,-} (f_i + f_i^\dagger) \, ,
\end{equation}
where $\gamma_{i,\pm} = (1 \pm \sigma_{i,z})/2$ and $\sigma_{i,\pm} =
(\sigma_{i,x} \pm i \sigma_{i,y})/2$. The spinless fermion operators
fulfill the canonical anticommutation relations
$\{f_i^\dagger,f_j\} = \delta_{i,j}$,
and $\sigma_{i,a}\, , \; a = x,y,$ or $z$, are the Pauli matrices. The 
Hamiltonian becomes
\begin{equation}
\tilde H_{t-J}= +t \sum\limits_{<i,j>} P_{ij}f^{\dagger}_if^{}_j +
\frac J 2 \sum\limits_{<i,j>}\Delta_{ij}(P_{ij}-1) ,
\label{glg:antimo}
\end{equation}
where
$P_{ij}=(1+\vec\sigma_i\cdot\vec\sigma_j)/2$,
$\Delta_{ij}=(1-n_i-n_j)$ and
$n_i=f^{\dagger}_if^{}_i$.
The constraint to avoid doubly occupied states transforms to the
conserved and holonomic constraint
$ \sum_i \gamma_{i,-} f^\dagger_i f_i = 0 $.
This constraint simply means, that a spinless fermion and a pseudospin 
$\downarrow$ are not allowed to sit on the same site.

In order to obtain the dynamics of the hole, we calculate the 
one-particle Green's function for spin up,
\begin{equation}
G (i-j, \tau)=
-\langle T \tilde c^{}_{i,\uparrow}(\tau) \tilde c^{\dagger}_{j,\uparrow}
\rangle
=
-\langle
 T f^{\dagger}_i(\tau) f^{}_j
\rangle
\end{equation}
where $T$ corresponds to the time ordering operator.
Inserting complete sets of spin states the quantity above
transforms as
\end{multicols}
\begin{eqnarray}
G(i-j,-\tau)
&=&
\frac{\sum \limits_{\sigma_1}
\langle v | \otimes \langle \sigma_1 |
e^{-(\beta-\tau)\tilde H_{t-J}}f^{}_j
e^{-\tau \tilde H_{t-J}}f^{\dagger}_i |
\sigma_1 \rangle \otimes | v \rangle}
{\sum \limits_{\sigma_1}\langle \sigma_1 |
 e^{-\beta\tilde H_{t-J}} | \sigma_1 \rangle}
  \nonumber\\
&=& \sum\limits_{\vec\sigma} P(\vec\sigma) \times
\frac{
\langle v|  f^{}_j
e^{-\Delta\tau\tilde H(\sigma_n,\sigma_{n-1})}
\ldots
 e^{-\Delta\tau\tilde H(\sigma_{2},\sigma_{1})}
 f^{\dagger}_i
|v \rangle
}
{
\langle\sigma_n |  e^{-\Delta\tau\tilde H_{t-J}} | \sigma_{n-1}\rangle
\ldots \langle \sigma_{2}|  e^{-\Delta\tau\tilde H_{t-J}}
| \sigma_{1}\rangle
}
+{\cal O}(\Delta\tau^2)
 \nonumber\\
&=&
\sum\limits_{\vec\sigma} P(\vec\sigma) G(i,j,\tau,\vec\sigma)
+{\cal O}(\Delta\tau^2)
\label{glg:pgr}
\end{eqnarray}
\begin{table}[bt]
\begin{tabular}{|c|c|c|}
${ x}\rightarrow { x}$ & ${ x}+\vec \delta \rightarrow { x}$ &
spin configuration \\ \hline
$0$ & $0$ &\mbox{\epsfig{file=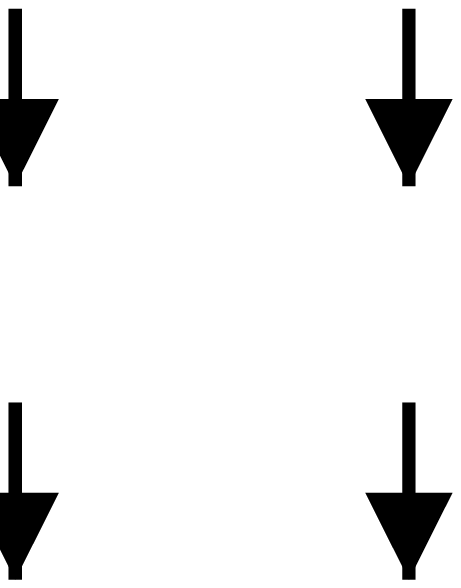,height=0.42cm}}\\ \hline
$\cosh(\Delta\tau t)$ & $-\sinh(\Delta\tau t)$ &
\mbox{\epsfig{file=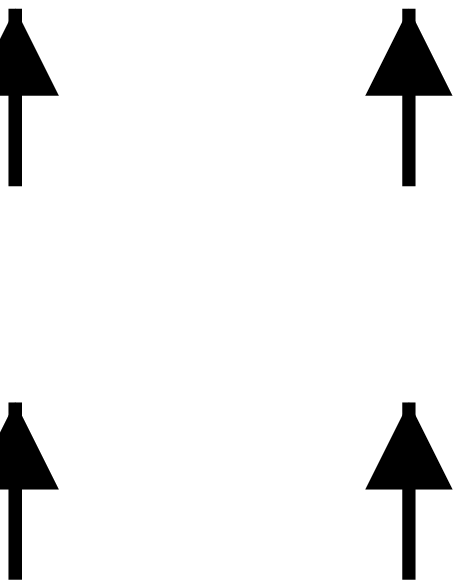,height=0.42cm}}\\ \hline
$\frac{\cosh(\Delta\tau t)}{\exp(\Delta\tau J/2)\cosh(\Delta\tau J/2)}$ &
$0$ &
\mbox{\epsfig{file=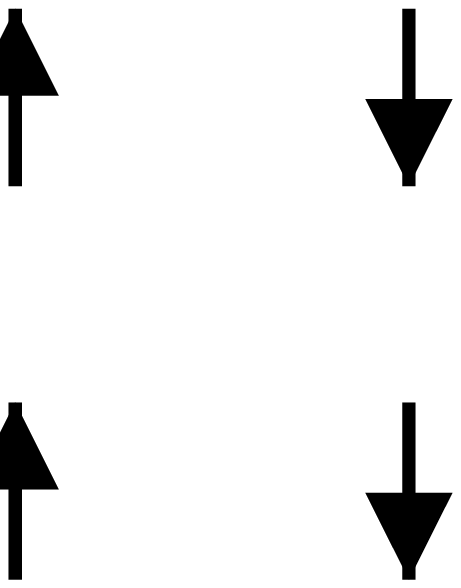,height=0.42cm}}\,,\,\,
\mbox{\epsfig{file=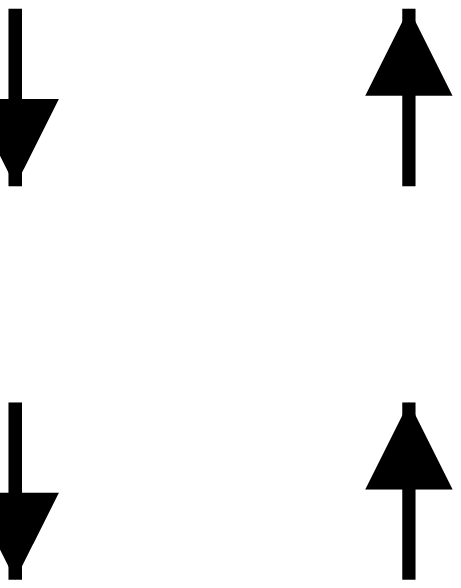,height=0.42cm}}
\\ \hline
$0$ & $\frac{\sinh(\Delta\tau t)}{\exp(\Delta\tau J/2)
\sinh(\Delta\tau J/2)}$ &
\mbox{\epsfig{file=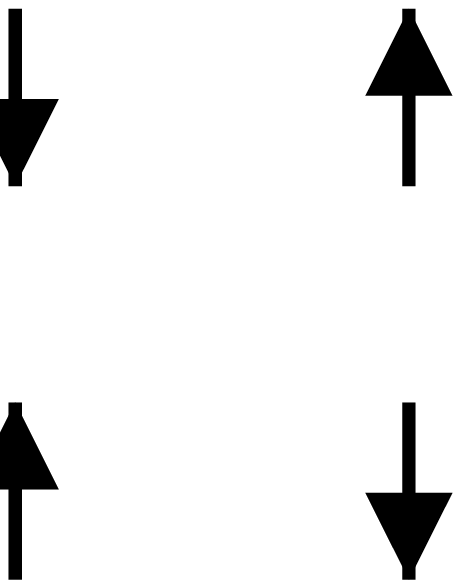,height=0.42cm}}\,,\,\,
\mbox{\epsfig{file=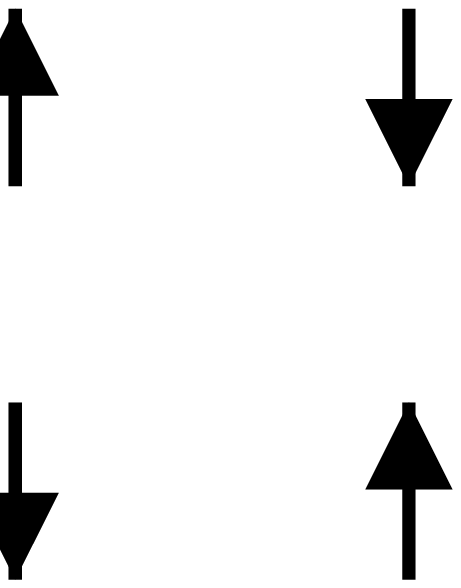,height=0.42cm}}
\end{tabular}
\caption{
Contributions for the propagation of the hole on one
plaquette. The first column shows the weight for a
propagation where the hole stays on
the same site $x$, whereas in the second column the weight
corresponds to the propagation
to the adjacent site.                                                                  
The third column represents the spin background on the plaquette. }                    
\label{table:prop}                                                                     
\end{table}\widetext
\begin{multicols}{2}

Here $m \Delta\tau=\beta$, $n \Delta\tau=\tau$, $\Delta\tau t \ll 1$
and $\exp\bigl({-\Delta\tau\tilde H(\sigma_1,\sigma_2)}\bigr)$ is the
evolution operator for the holes,
given the spin configuration $(\sigma_1,\sigma_2)$.
In the case of single hole dynamics,
$|v\rangle$ is the vacuum state for  holes, and
$P(\vec\sigma)$ is the probability distribution of
a Heisenberg antiferromagnet for the configuration $\vec\sigma$, where
$\vec \sigma$ is a vector containing all intermediate states
$(\sigma_1,\ldots\sigma_n,\ldots,\sigma_1)$.
The sum over spins is performed in a very efficient way
by using a world-line loop-algorithm \cite{evertz93} for a Heisenberg 
antiferromagnet with discretized imaginary time.
In general we have $\Delta\tau=0.05$, such that the extrapolation to 
$\Delta\tau=0$ leads to values of the observables
within the statistical error bars. The inverse temperature $\beta$ is
taken such that the energy is well converged ($\beta J \geq 20$ for 
$16 \times 16$ and $\beta J = 30$
for $24 \times 24$ sites), and therefore,
the data correspond to the ground-state.  
As the evolution operator for the holes is a bilinear form
in the fermion operators,
$G(i,j,\tau,\vec\sigma)$ can be calculated exactly, in contrast
to a direct implementation in the
loop algorithm \cite{prokofev98,brower98},
where fermion paths are sampled stochastically.
$G(i,j,\tau,\vec\sigma)$ contains a sum over all possible fermion paths
between $(i,0)$ and $(j,\tau)$. 
The numerical effort to calculate
$G(i,j,\tau,\vec\sigma) \, \forall \, i,\tau$  scales as $N \tau$,
where N is the number of lattice points in space. Therefore, the present
method is more efficient for large systems than
e.g.\ projector algorithms 
for the Hubbard model, that scale with the system size cubed.

With the representation of Eq.\ (\ref{glg:antimo}), the propagation of 
down spin electrons cannot be easily considered,
since the operators $\sigma_{i,\pm}$ cut world-lines.
This is certainly not a problem for finite-size systems,
where SU(2) symmetry is conserved.
Since $P(\vec\sigma)$ is the probability distribution
for the quantum antiferromagnet, the algorithm does not suffer from
sign problems
on bipartite lattices and
non-frustrating magnetic
interactions in any dimension.

We now address the explicit calculation of $G(i,j,\tau,\sigma)$.
In a first step, we introduce additional complete sets of single fermion
states in the numerator of Eq.(\ref{glg:pgr}),
such that $G(i,j,\tau,\vec\sigma)$ becomes 
\end{multicols}
\begin{eqnarray}
&&
\bra{v}f^{}_j
\Bigl(
\sum\limits_{\vec l}
\ket{f_{l_n}}
\frac{\bra{f_{l_n}}
e^{-\Delta\tau\tilde H(\sigma_n,\sigma_{n-1})}
\ket{f_{l_{n-1}}}}
{\bra{\sigma_{n}}
e^{-\Delta\tau H}\ket{\sigma_{n-1}}}
\frac{\bra{f_{l_{n-1}}}
e^{-\Delta\tau\tilde H(\sigma_{n-1},\sigma_{n-2})}
\ket{f_{l_{n-2}}}}
{\bra{\sigma_{n-1}}
e^{-\Delta\tau H}\ket{\sigma_{n-2}}}
\ldots\nonumber\\
&&
\times \ldots
\frac{\bra{f_{l_2}}
e^{-\Delta\tau\tilde H(\sigma_2,\sigma_1)}
\ket{f_{l_{1}}}}
{\bra{\sigma_{2}}
e^{-\Delta\tau H}\ket{\sigma_{1}}}
\bra{f_{l_1}}
\Bigr)
f^{\dagger}_i\ket{v}\nonumber\\
&&=
\sum\limits_{\vec l}\bigl(
\delta_j^{l_n} U(\sigma_{n},\sigma_{n-1})_{l_{n}}^{l_{n-1}}\ldots
U(\sigma_2,\sigma_1)_{l_2}^{l_1}    \delta^i_{l_1}\bigr),
\label{glg:matr}
\end{eqnarray}
\begin{multicols}{2}
where the sum $\sum_{\vec l}$ runs over all possible intermediate one 
particle states in the fermionic Hilbert space $\{\ket{f_l}\}$.
The propagators
${\bra{f_{l_p}}
e^{-\Delta\tau\tilde H(\sigma_p,\sigma_{p-1})}
\ket{f_{l_{p-1}}}}/
{\bra{\sigma_{p}}e^{-\Delta\tau H}\ket{\sigma_{p-1}}}
$
are only nonzero, when $l_p$ and $l_{p-1}$ belong to the same plaquette.
Therefore the entries of the matrices $U(\sigma_p,\sigma_{p-1})$ are 
nonzero only at the positions which correspond to a plaquette in the 
checkerboard breakup. These entries are given in Table~\ref{table:prop}.
As we are only interested in the Hilbert space of no double occupancy,
we have to enforce the
constraint at one single position of the propagation
by projecting out the fermionic states
which do not respect the constraint.
We do so at $\tau=0$ corresponding to the first propagation.

The one-particle spectral function
\begin{equation}
A(\vec k,\omega)
=\sum\limits_{f,\sigma}
\left|
\bra{f,N-1}c^{}_{\vec k,\sigma}
\ket{0,N}
\right|^2
\delta\left(
\omega-E_0^N+E_f^{N-1}\right),
\end{equation}
is connected with the Green's function in imaginary time at $T=0$,
by the spectral theorem
\begin{equation}
G(\vec k,\tau)=
\int\limits_{-\infty}^{\infty}
d\omega \frac {\exp(-\tau\omega)} {\pi} A(\vec k,\omega)     
\label{glg:STT0}.
\end{equation}
Here $\ket{0,N}$ is the ground-state at half filling with energy $E_0^N$
and $\ket{f,N-1}$ are states in the $N-1$ particle Hilbert space 
with energy $E_f^{N-1}$.
We perform the inversion of Eq.~(\ref{glg:STT0}), that due to the 
statistical errors of $G(\vec k,\tau)$ is an extremely ill-posed problem,
by means of MaxEnt, where the 
$A(\vec k,\omega)$ obtained is the one that 
maximizes the probability $P(A|G)$, given
the Green's function $G(\vec k,\tau)$.
Correlations in the imaginary time data were taken into account by
considering the covariance matrix.
Details about MaxEnt can
be found in the comprehensive review article by J.E.\ Gubernatis and 
M.\ Jarrell \cite{jarrell96}.

We would like to stress finally, that part of the dynamical data presented
below were obtained without use of MaxEnt but directly extracted from 
the imaginary time Green's function. This is possible due to the high 
statistics and stability attainable with the present algorithm. The 
slowest decaying exponential, that corresponds to the excitation with
lowest energy can be extracted
simply by fitting the tail of the Green's function at large values
of $\tau$. This leads to the value of the excitation and its corresponding
weight, as shown in Secs.\ \ref{sec:low} and \ref{sec:qp}. 
Furthermore, in connection with MaxEnt,
the next higher excitation can be obtained by subtracting the contribution
from the lowest one from the Green's function. This procedure is discussed
in Sec.\ \ref{sec:sf}.
\section{Results}
\label{sec:res}
We concentrate in the following on three aspects
of the dynamics of a single 
hole in a Heisenberg antiferromagnet. First we consider in Sec.\ 
\ref{sec:low} the lower edge of the spectrum. 
This is a quantity that can be obtained by several other methods,
including various Monte Carlo algorithms, such that the relative accuracy 
of each one and the region in parameter space,
where each method gives best 
results, can be assessed. In our case, this quantity is obtained from the
asymptotic behavior of the one-particle Green's function
in imaginary time.
However, not only the energy but also the weight of such an excitation can
be extracted from the asymptotics, leading to the quasiparticle weight, as
discussed in Sec.\ \ref{sec:qp}. The present algorithm 
is up to now the only one capable of extracting this information 
for the $t$-$J$ model free of approximations
on large lattices (in general up to $16 \times 16$ and for $J/t = 2$
up to $24 \times 24$ sites). 
For small lattice sizes, the results can be compared with 
exact diagonalizations, whereas for 
large systems only comparisons with
approximate methods like SCBA can be made.  
Finally, the whole spectrum is considered in Sec.\ \ref{sec:sf}, where
the spectral function $A (\vec k , \omega)$ 
is discussed. Using the information from the lower edge of the spectrum, 
a resonance above the quasiparticle band 
is identified, that is very well described as 
a string excitation. 
%
\begin{figure}[tb]\narrowtext
\noindent \epsfig{file=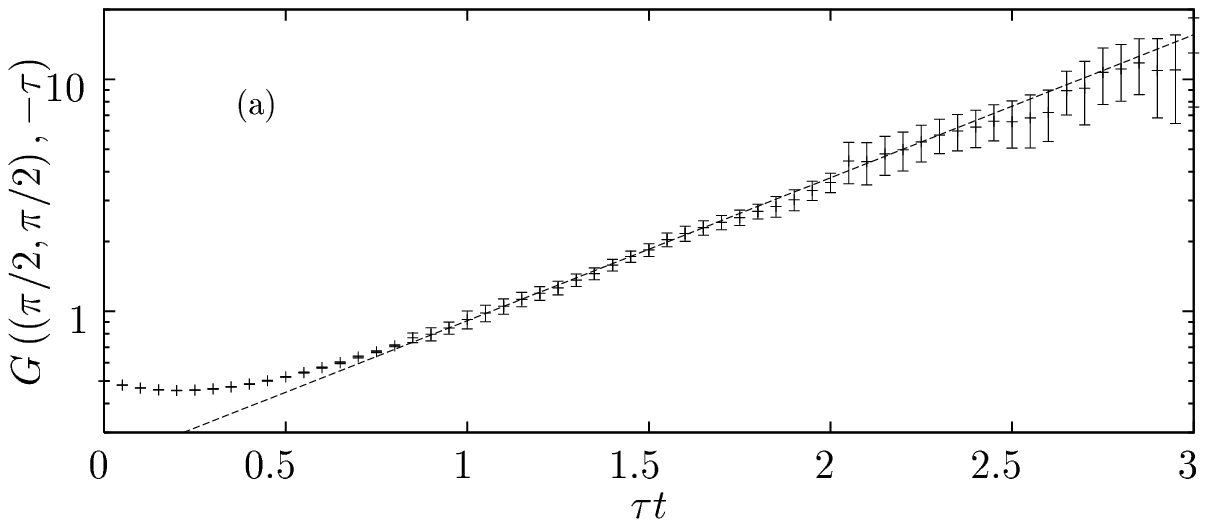,width=8cm}
\epsfig{file=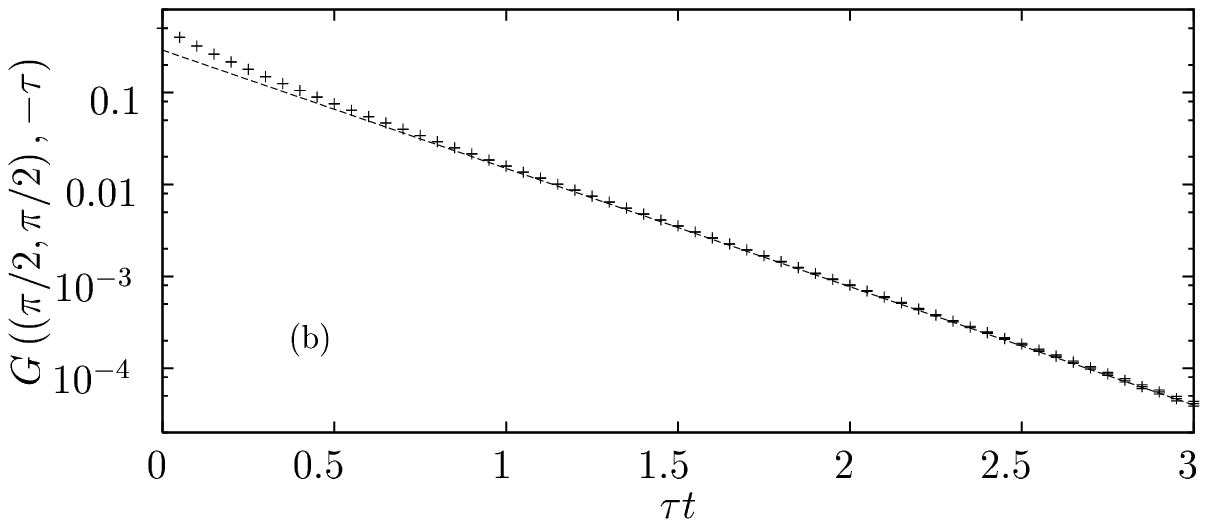,width=8cm}
\caption{\label{fig:Elt}
The energy of the lowest excitation is extracted from the imaginary
time asymptotics of the Green's function, as indicated by the
dotted line for a) $J=0.4 t$ and b) $J = 2 t$ in a $16\times 16$ lattice.}
\end{figure}

\subsection{The lower edge of the spectrum}
\label{sec:low}
The accuracy and stability of the data allow in our case to obtain the 
lower edge of the spectrum directly 
from the slope of the one-particle Green's function as a function of
imaginary time $\tau$, for large values of $\tau$.
Figure \ref{fig:Elt} shows the 
asymptotics in imaginary time for two values of the coupling constant,
showing that the most accurate results are obtained, when $J/t = 2$.
$J/t=0.4$ is the smallest coupling, where such a procedure can be applied.
In order to check the results obtained at the smallest coupling,
we made additional calculations at $\Delta\tau t=0.2$
(all other calculations are done at $\Delta\tau t=0.05$),
where larger values of $\tau t$ can be reached.
The resulting Green's functions are the same within the error bars,
indicating a small $\Delta\tau$ effect. 

\begin{figure}[tb]\narrowtext
\noindent \epsfig{file=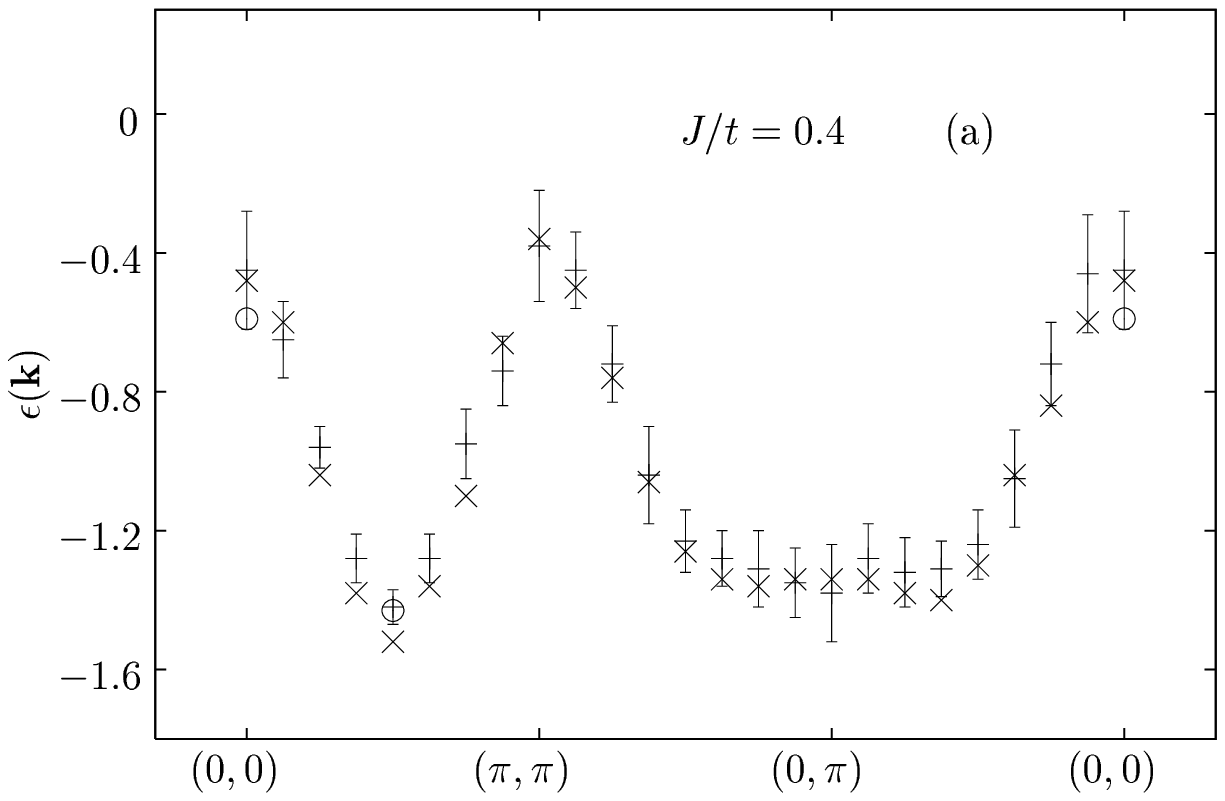,width=8cm}
\epsfig{file=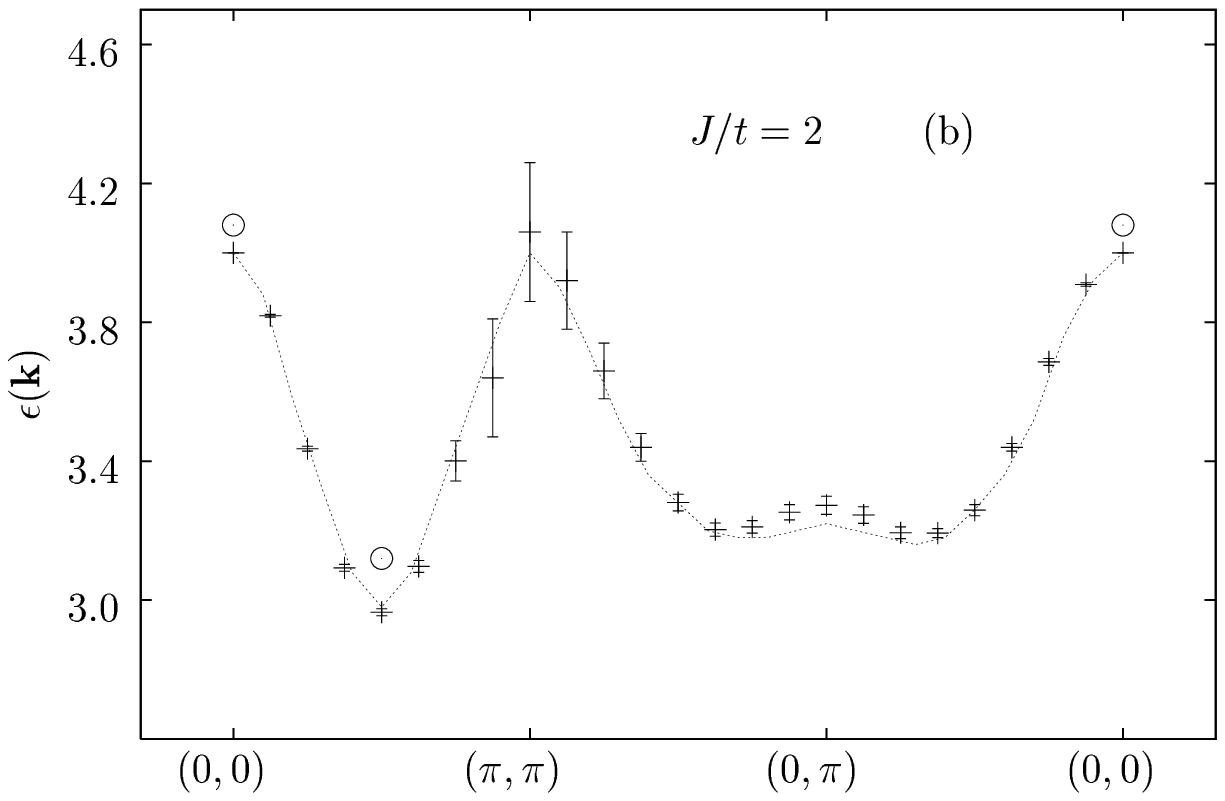,width=8cm}
\caption{ Lower edge of the spectrum along the symmetry
lines of the Brillouin zone for a) $J/t=0.4$ and b) $J/t=2$
in a $16\times 16$ lattice.
Comparisons are made with VMC (circles), GFMC for $J/t=0.4$ ($\times$),
and series expansion \cite{hamer98} for $J/t=2$ (dotted line).
\label{fig:2DGF}}
\end{figure}

Figure \ref{fig:2DGF} shows the lower edge of the spectrum for 
$J/t =0.4$ and $J/t=2 $ in a $16 \times 16$ sites system.
The energies are displayed with respect to 
the ground-state energy of the Heisenberg antiferromagnet. 
The results are compared with 
variational Monte Carlo (VMC) \cite{boninsegni92}, Green's function
Monte Carlo (GFMC) \cite{boninsegni94}, and series expansions
\cite{hamer98}, whenever data is available. At $J/t=0.4$ 
(Fig.\ \ref{fig:2DGF}a), where our results
are most affected by fluctuations, we observe good agreement with GFMC.
The behavior of the statistical error is similar in both methods, with 
larger fluctuations around $\vec k = (0,0)$ and $(\pi,\pi)$. Around 
$\vec k = (\pi,0)$ our results show somewhat larger fluctuations. 
For $J/t = 0.4$ VMC \cite{boninsegni92}, also appears to be 
very accurate concerning the lower edge.
When its energies are  compared  to
our calculations and the GFMC technique, we find that their
energies are within the error bars of the exact QMC calculations.
At ${\vec k}=(0,0)$, the variational result is
at the lower edge of the error bars of our calculation,
and have the smallest statistical error of all three approaches.
At this specific $k$-point both GFMC and our approach 
have large fluctuations before the state with lowest energy is clearly 
reached. As mentioned above, 
additional calculations with $\Delta\tau t=0.2$ were performed, in order
to check the results obtained, without observing significant changes.

\begin{figure}[tb]
\noindent\epsfig{file=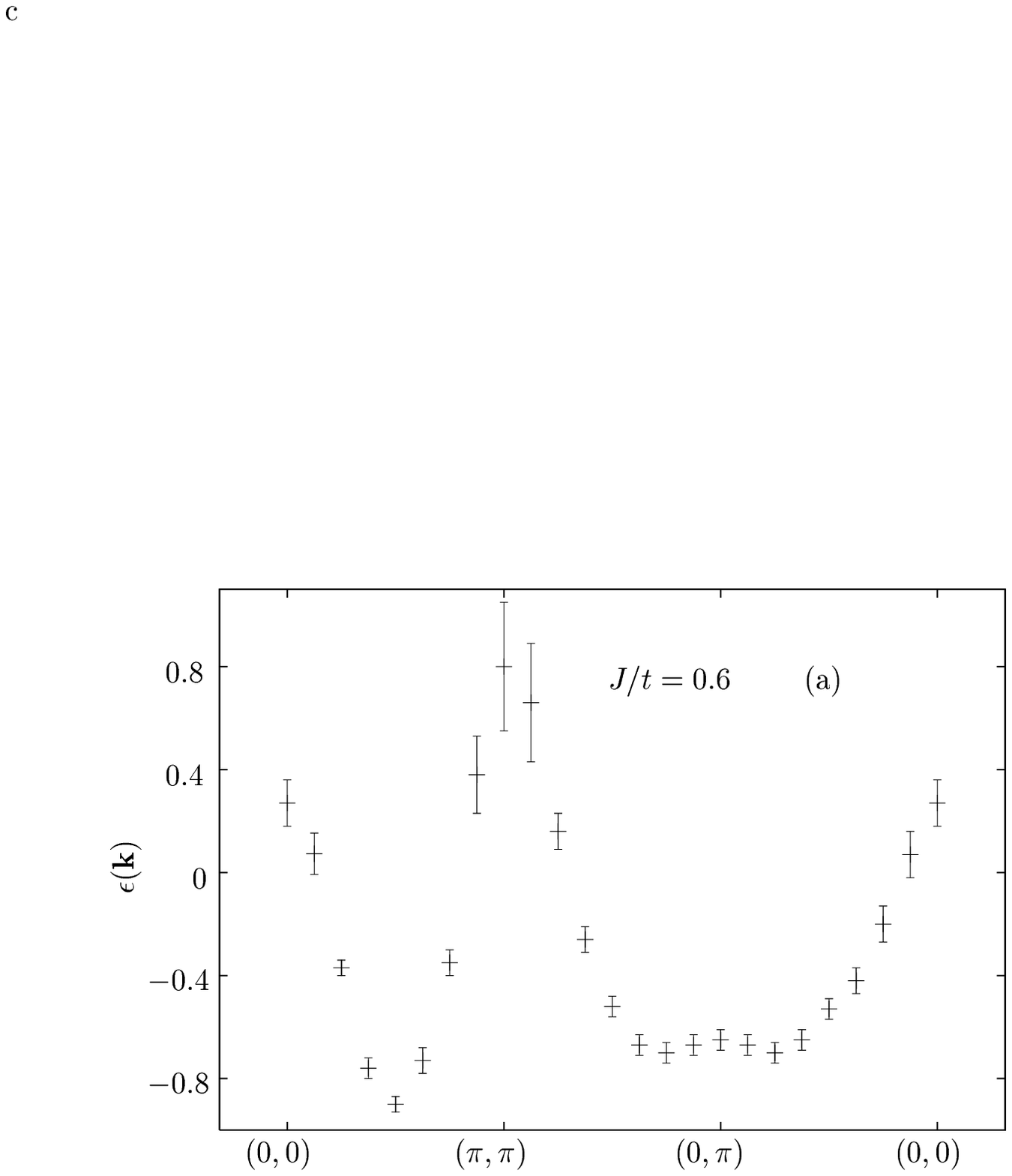,width=8cm}
\epsfig{file=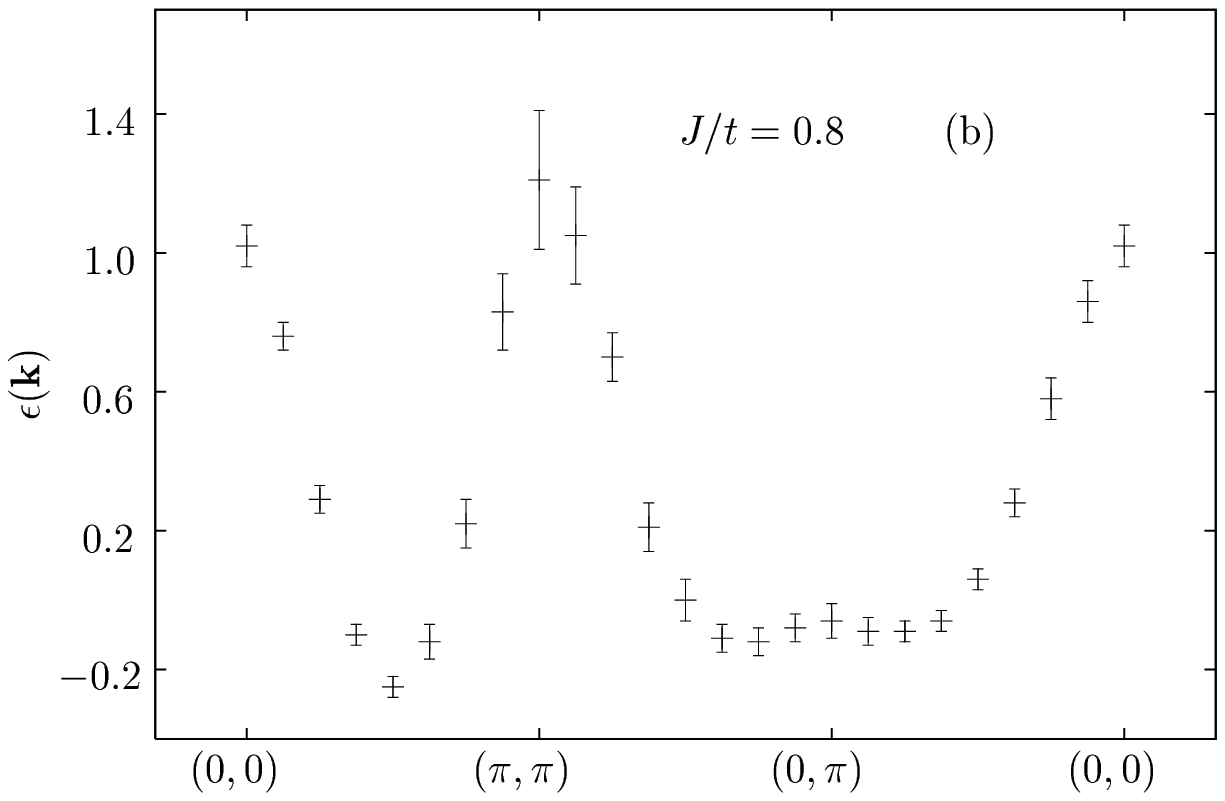,width=8cm}

\noindent
\epsfig{file=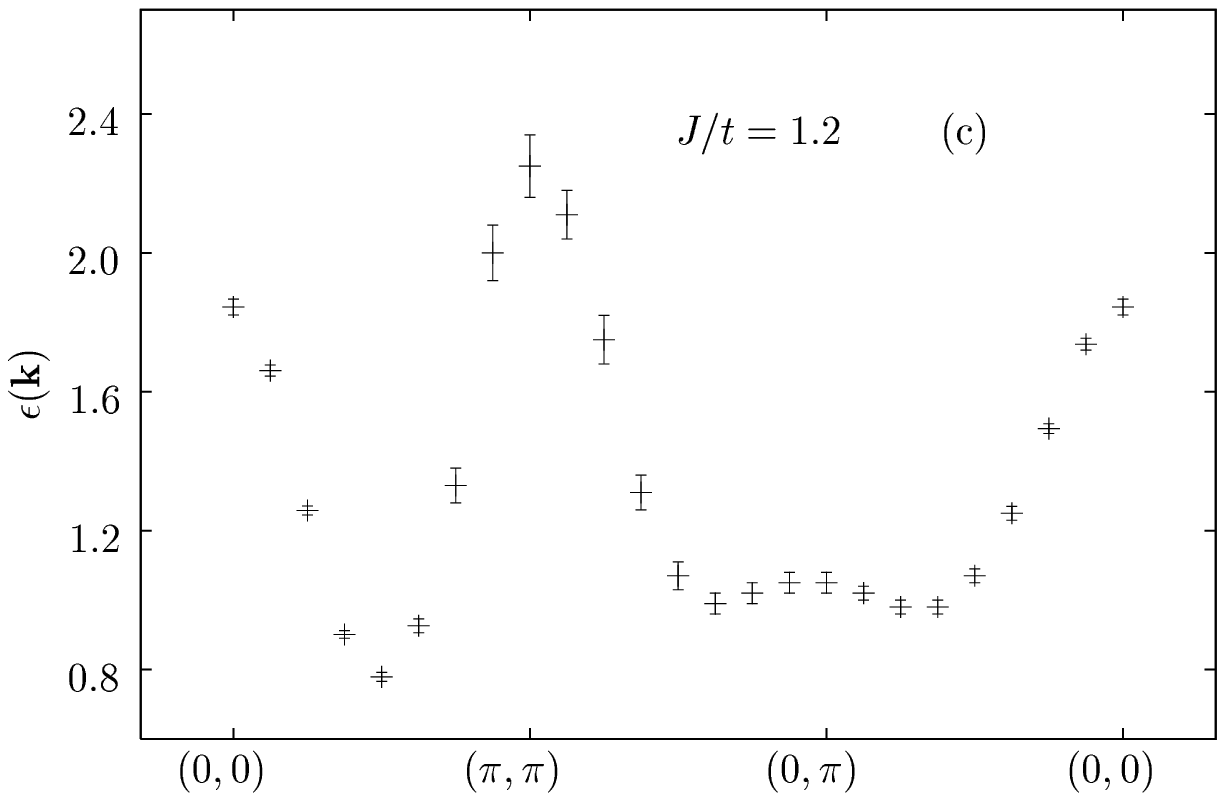,width=8cm}
\epsfig{file=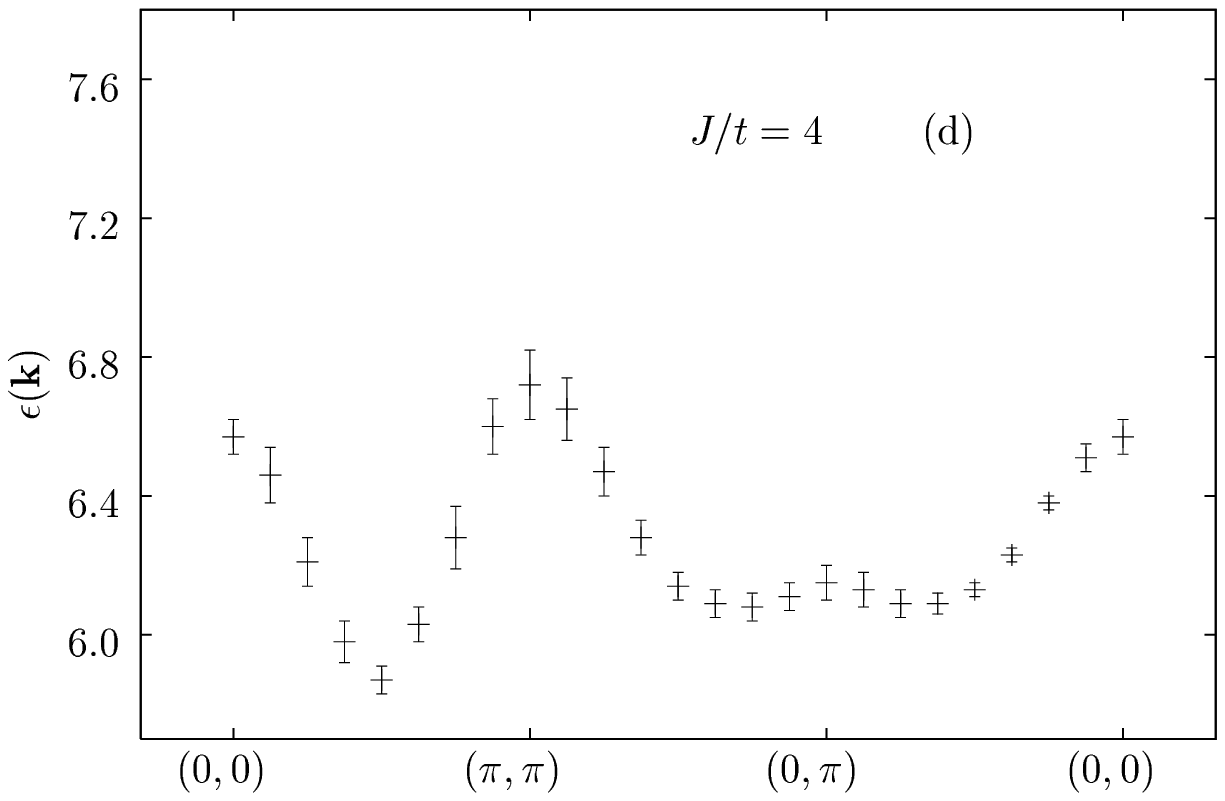,width=8cm}
\caption{\narrowtext Lower edge of the spectrum along the symmetry
lines of the Brillouin zone for a) $J/t=0.6$, b) $J/t=0.8$,
c) $J/t=1.2$, and d) $J/t=4$ in a $16\times 16$ lattice.
\label{fig:2DGFm}}
\end{figure}

Figure \ref{fig:2DGF}b shows that 
at $J/t=2$, where our algorithm leads to much more accurate results,
the variational results are too high 
in energy, but still close to our numerically exact ones.
For values of $J/t \ge 1$, additional results from 
series expansions \cite{hamer98} are available. At $J/t=2$ we observe 
in general a very good agreement. Only around $(\pi,0)$
we see that series expansions slightly underestimate the
energy of the hole. 
The general features of the lower edge are not substantially modified
when going from $J/t = 0.4$ to $J/t =4$. This is shown in 
Fig.\ \ref{fig:2DGFm},
where the only changes observed are an overall shift
in energy with respect 
to the Heisenberg antiferromagnet and a change in the bandwidth.
The shift in energy can be 
followed by considering the dependence of $\epsilon(\pi/2,\pi/2)$ on 
$J/t$. This dependence is rather accurately
described by $\epsilon(J/t)/t = -3.28 + a_1 (J/t)^{2/3}$, where $a_1$
is the first eigenvalue of the dimensionless Airy equation
(see Fig.\ \ref{fig:3exJ06} in Sec.\ \ref{sec:sf}). Such a scaling
of the hole energies is found in the $t$-$J_z$ model
in the continuum limit
for small values of $J_z$ \cite{bulaevskii68,kane89,liu92},
when loops along the path of the hole are disregarded.
In that case, the constant is 
$-2 \sqrt{z-1}$, where $z$ is the coordination number. 
The resulting string picture gives an
accurate description of the lowest excitations close to 
$\vec k = (\pi/2,\pi/2)$.
As will be shown in Sec.\ \ref{sec:sf}, also the 
next higher excitation can be described by the string picture.       

\begin{figure}[htb]
\noindent \epsfig{file=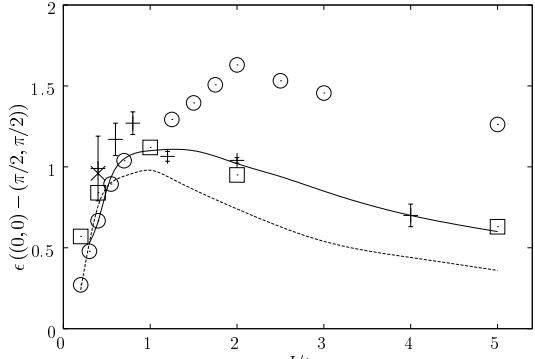,width=8cm}
\caption{\narrowtext Bandwidth of the lower edge as a function of $J/t$.
$\epsilon(0,0)-\epsilon(\pi/2,\pi/2)$
in a $16\times 16$ lattice compared with exact diagonalization
($4 \times 4$ sites, open circles), GFMC (cross),
VMC (open boxes), SCBA (dashed line), and series expansions (full line).
\label{fig:2DBW}}
\end{figure}

Figure \ref{fig:2DBW} shows the bandwidth obtained in our simulations 
compared with exact diagonalizations \cite{dagotto90},
GFMC \cite{boninsegni94},
SCBA \cite{martinez91,liu92}, VMC\cite{boninsegni92} and 
series expansions \cite{hamer98}. 
For $J/t < 0.8$ good agreement is found among all methods,
whereas for larger
values of $J$, only series expansions and VMC agree with our data. 
This, and the fact that the string picture gives a good representation 
of the lowest lying states, suggest that a perturbation expansion 
as performed in series expansions can 
be used to interpret the distinctive features of the lower edge. 
In particular the flat band observed around $\vec k = (\pi,0)$ and the 
fact that the degeneracy between this point and $\vec k = (\pi/2,\pi/2)$ 
suggested by some approaches \cite{eder90} is clearly lifted, as shown
by our simulation, are very well reproduced by series expansions.
The flat bands can be well observed for all considered values of
$J/t$, when considering the lower edge~(Fig.~\ref{fig:2DGF}
and~\ref{fig:2DGFm})
and the complete spectral function~(Fig.~\ref{fig:Ak2DJ04}).
Our data clearly show for $J/t\ge 0.6$, that the neighboring points of
${\vec k}=(\pi,0)$ are generally slightly higher in energy.
The band in this area does not seem to be completely flat,
but it changes its
curvature with local minima of the dispersion at
the points $(\pi,\delta)$ and $(\pi-\delta,0)$, when going along
the (1,0) and (0,1) directions respectively, with the {\em caveat} that
they are well defined beyond the error bars only for $J/t > 1$. In all
the cases we find $\delta\approx 0.3\pi$. This region
with a very flat band  
spans an extremely large area in the Brillouin zone.

A flat band on a similarly wide region in the Brillouin 
zone around ${\vec k}=(\pi,0)$
is also observed in photoemission spectroscopy 
of cuprates close to the Fermi-energy in the optimally doped compounds.
As doping is reduced, that portion of the spectrum opens a pseudogap 
and weight is transferred to higher energies \cite{marshall96},
until in the 
undoped materials, this portion is about $2 J$ ($\approx 300$meV)
above the minimum  
at ${\vec k}=(\pi/2,\pi/2)$ \cite{wells95,Kim98,Ronning98}.
The energy difference between the points ${\vec k}=(\pi/2,\pi/2)$
and  ${\vec k}=(\pi,0)$ is in our simulation
about $\Delta=(0.25  \pm 0.10) t$ ($\approx J/2$ for $J=0.4t$).
The rather large error corresponds
mainly to $J/t < 1$. No significant
dependence on $J/t$ can be observed in the whole range under
consideration, 
in contrast to the results from SCBA and series expansions. However, 
it could be that the $J$-dependence 
is masked in our case by large fluctuations, taking into 
account that the variations observed for this quantity 
by SCBA and series expansions 
are much smaller than the one observed for the bandwidth.   
SCBA \cite{martinez91} gives values ranging from
$0.17 t$ ($J/t=1$) to $0.12 t$ ($J/t=4$), that are smaller
than the values we obtain. On the other hand,
series expansions \cite{hamer98} obtain values
between $0.15 t$ at $J/t=1$ and $0.25 t$ at $J/t= 2.5$.
The values obtained by series expansions are consistent with our results
for large values of $J/t$.

\subsection{The quasiparticle weight}
\label{sec:qp}
The quasiparticle weight is the weight of the 
exponential with the slowest decay, that is the exponential 
that determines the lower edge of the spectrum. This weight
is
\be
\label{glg:qp}
Z(\vec k) = \lim_{-\tau \rightarrow \infty}
G (-\tau,\vec k) \exp \left[ \left(\epsilon_{\vec k}
- \epsilon_0 \right) \tau \right]
\ee

\begin{figure}[ht]
\noindent \epsfig{file=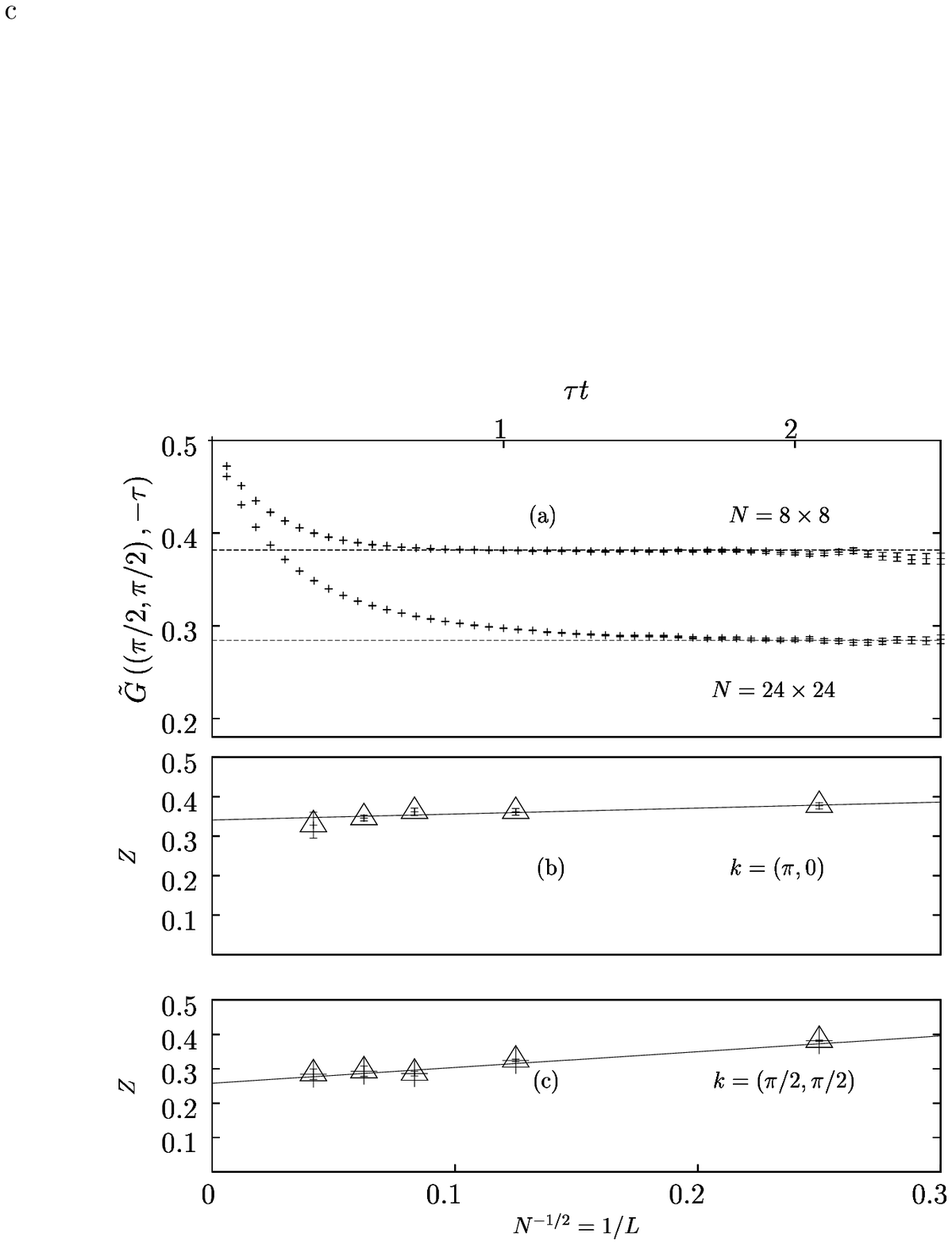,width=8cm}
\caption{\narrowtext a) Extrapolation of $\tilde G(\vec k,-\tau) \equiv
G(\vec k, -\tau) \exp[(\epsilon_k - \epsilon_0) \tau]$ for
$N = 8 \times 8$ and $24 \times 24$ at $J/t =2$.
Finite-size scaling for b) ${\vec k}=(\pi/2,\pi/2)$
and c) ${\vec k}=(\pi,0)$.\label{fig:Z2DJ2}}
\end{figure}

\begin{figure}[tb]                                                                    
\noindent \epsfig{file=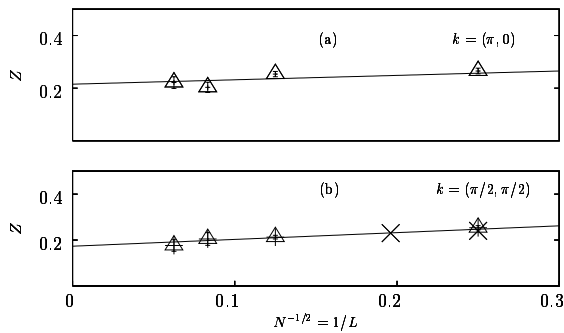,width=8cm}                                                 
\caption{\narrowtext Finite size scaling of $Z (\vec k)$                                           
at $J/t=0.6$ for a) ${\vec k}=(\pi/2,\pi/2)$                                           
and b) ${\vec k}=(\pi,0)$. The crosses are values                                      
from exact diagonalization results \cite{dagotto90,beran96}.                           
\label{fig:Z2DJ06}}                                                                    
\end{figure}

In the following we focus on the thermodynamic limit of $Z(\vec k)$
for the wave vectors
${\vec k}=(\pi,0)$ and ${\vec k}=(\pi/2,\pi/2)$.
Figures \ref{fig:Z2DJ2} and \ref{fig:Z2DJ06}
show the finite-size scaling on these two points for $J/t =2$ and
$J/t = 0.6$ respectively. For both $k$- and $J$-values,
an appreciable quasiparticle weight is obtained, demonstrating that the
lower edge of the spectrum describes the band of a coherent quasiparticle.
The determination of the quasiparticle weight is only accurate for
$J/t \ge 0.6$. Below that value, the quality of the data is less 
satisfactory and, for $J/t=0.4$ 
the value presented can be taken only as an upper bound.
The size dependence of $Z(\pi/2,\pi/2)$ and $Z(\pi,0)$ is not very large 
and scales linearly with the inverse linear size of the system for 
$J/t \ge 0.6$, in agreement
with SCBA \cite{martinez91}. The size dependence at $(\pi/2,\pi/2)$ is 
systematically larger than  at $(\pi,0)$.
The sizes considered are $L \times L$, with $L = 16,12,8$, and 4.
At $J/t=2$ we use additionally a $24 \times 24$ lattice.
Values from exact diagonalization \cite{beran96,poilblanc93,dagotto90}
were included when available.

\begin{figure}[tb]                                                                    
\noindent\epsfig{file=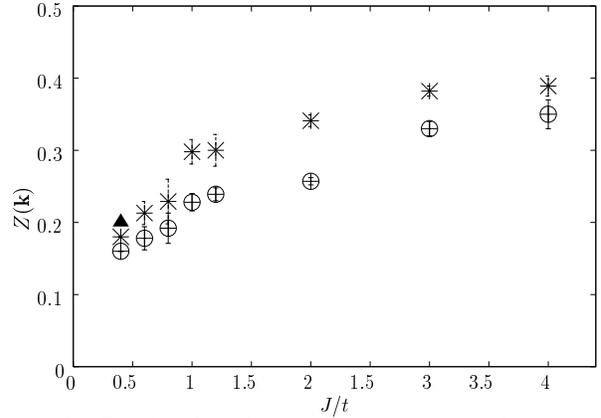,width=8cm}                                                  
\caption{\narrowtext Quasiparticle weight as a function of $J/t$ for                               
${\vec k}=(\pi/2,\pi/2)$ (circles) and ${\vec k}=(\pi,0)$ ($\times$).                  
The result from exact diagonalization \cite{dagotto90} in a $4 \times 4$               
lattice is given by the triangle.                                                      
\label{fig:ZJFS}}                                                                      
\end{figure}                   

\begin{figure}[tb]                                                                    
\noindent\epsfig{file=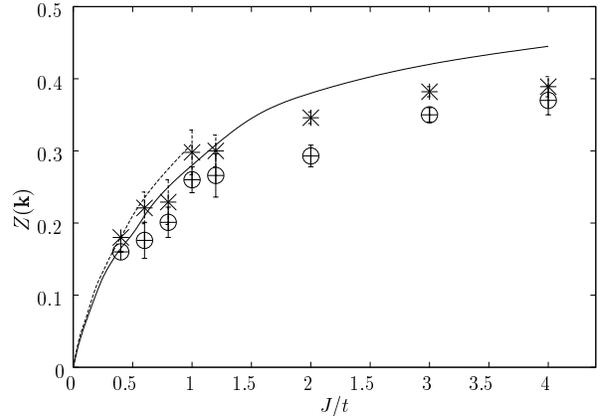,width=8cm}                                                  
\caption{\narrowtext Quasiparticle weight as a function of $J/t$ for                               
${\vec k}=(\pi/2,\pi/2)$ (circles) and ${\vec k}=(\pi,0)$ ($\times$) in                
a $16 \times 16$ lattice.                                                              
We compare our result with SCBA, where the dashed line corresponds to the              
quasiparticle weight for ${\vec k}=(\pi,0)$                                            
and the full line corresponds to                                                       
${\vec k}=(\pi/2,\pi/2)$. The data points were taken from                              
Ref.~11.                                                                               
\label{fig:ZJ16SCBA}}                                                                  
\end{figure}                         

Figure \ref{fig:ZJFS} shows that the extrapolated quasiparticle weight 
increases with $J/t$ both
for ${\vec k}=(\pi,0)$ and ${\vec k}=(\pi/2,\pi/2)$.
At $J/t=4$ the quasiparticle reaches about $80\%$ of its maximal value.
The changes of the quasiparticle weight with $J/t$
are small when $J/t\ge 1$ and the slope becomes 
steeper for smaller values. 
Estimates of the quasiparticle weight were given both by 
VMC \cite{boninsegni92} and SCBA~\cite{martinez91}, the difference 
being rather small. The general trend is that VMC
overestimates it at small
$J$ whereas SCBA overestimates it at large $J$.
For definiteness we compare our results with
SCBA for a $16 \times 16$ system in Fig.~\ref{fig:ZJ16SCBA}.
We find a rather good agreement between both methods. As in our case
$Z(\pi,0)>Z(\pi/2,\pi/2)$ for all considered values of $J/t$.
At small values of $J$ ($0.01\le J/t \le 0.5$) SCBA finds a scaling of
$Z(\pi/2,\pi/2)=0.31 J^{2/3}$ and $Z(\pi,0)=0.35 J^{0.7}$.
For $J/t \ge 1$, the results from SCBA overestimates the
quasiparticle weight at the two considered $k$-points, with an
increasing deviation for larger values of $J/t$. 
Based on the quantitative agreement of SCBA with
our results for small $J$, we can confidently conclude that  
the quasiparticle at ${\vec k}=(0,\pi)$ and $(\pi/2,\pi/2)$
should be finite for all values of $J$ in the physically relevant region
(i.e.\ $J/t \gtsim 0.1$).

\begin{figure}[tb]                                                                    
\noindent\epsfig{file=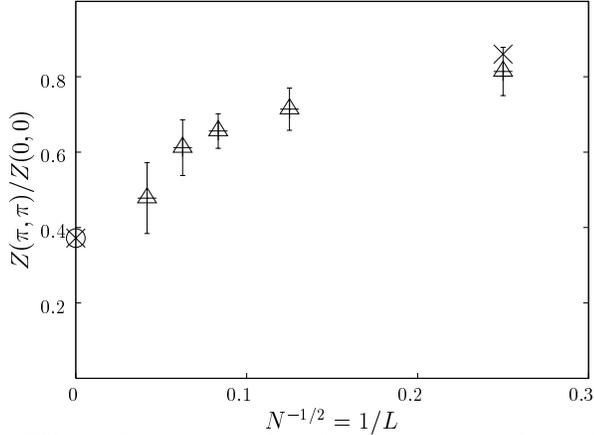,width=8cm}                                                  
\caption{\narrowtext Finite size scaling for the quasiparticle weight at                           
$\vec Q = (\pi,\pi)$ for $J/t = 2$. The cross in the thermodynamic limit               
is $(2m)^2$, $m$ being the staggered magnetization.                                    
\label{fig:Zpipi}}                                                                     
\end{figure}                                                                           
                                                                                       
\begin{figure}[tb]                                                                    
\noindent\epsfig{file=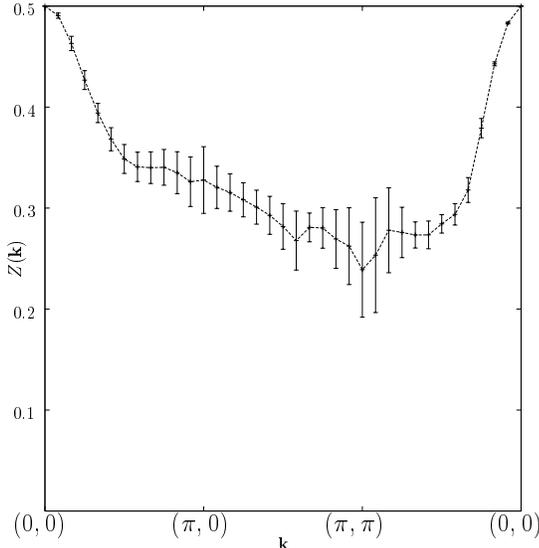,width=8cm}                                                 
\caption{\narrowtext $Z(\vec k)$ along the symmetry lines in the Brillouin zone                    
for $J/t = 2$ in a $24 \times 24$ lattice.                                             
\label{fig:ZJ2-24x24}}                                                                 
\end{figure}                         

As mentioned in the introduction, there are exact results for the 
quasiparticle weight at the supersymmetric point in two dimensions
\cite{sorella96b}. On the one hand, $Z(\vec k = 0) = 1/2$, a
requirement that is fulfilled by our simulation, where the Green's
function at that particular $k$-point consists of a single exponential.
In contrast to this, the estimate of SCBA is approximately $0.45$ and that
of VMC $\approx 0.32$. Furthermore, Sorella showed that 
$Z(\vec Q )/Z(0) \leq (2 m)^2$, where $m^2 = S(\vec Q )/N$, 
$S(\vec Q)$ being the magnetic structure factor at the antiferromagnetic
wave vector. The equality is reached in the thermodynamic limit. 
Figure \ref{fig:Zpipi} shows the evolution with system size of 
$Z(\vec Q )$ together
with results from exact diagonalization for a $4 \times 4$ system and
$(2m)^2 \simeq 0.37$ for $L \rightarrow \infty$.
Although large error bars 
show that the determination of $Z(\vec k)$ is less satisfactory for 
$\vec k = \vec Q$ than at $\vec k = (\pi/2,\pi/2)$,
the data are consistent
with the exact result. It was further suggested \cite{sorella96b} that if 
$Z(\vec k + \vec Q )/Z(\vec k ) = (2 m)^2$
is satisfied for $\vec k \neq 0$, a jump in the quasiparticle 
weight should be observed on crossing the border of the magnetic zone.
Figure \ref{fig:ZJ2-24x24} shows $Z(\vec k )$ along the symmetry 
directions in the 
Brillouin zone for a $24 \times 24$ system and $J/t = 2$. Our data
do not show any sizable jump. Unfortunately, it is not 
possible to consider arbitrarily long imaginary times since as Eq.\
(\ref{glg:qp}) shows, the errors are amplified exponentially. Therefore,
our results cannot be considered as a proof of continuity.
However, in view
of the good agreement with the above mentioned exact results, we consider
them as a convincing evidence. 
\subsection{Spectral function and string excitations}
\label{sec:sf}

\begin{figure}[hbt]                                                                    
\noindent\epsfig{file=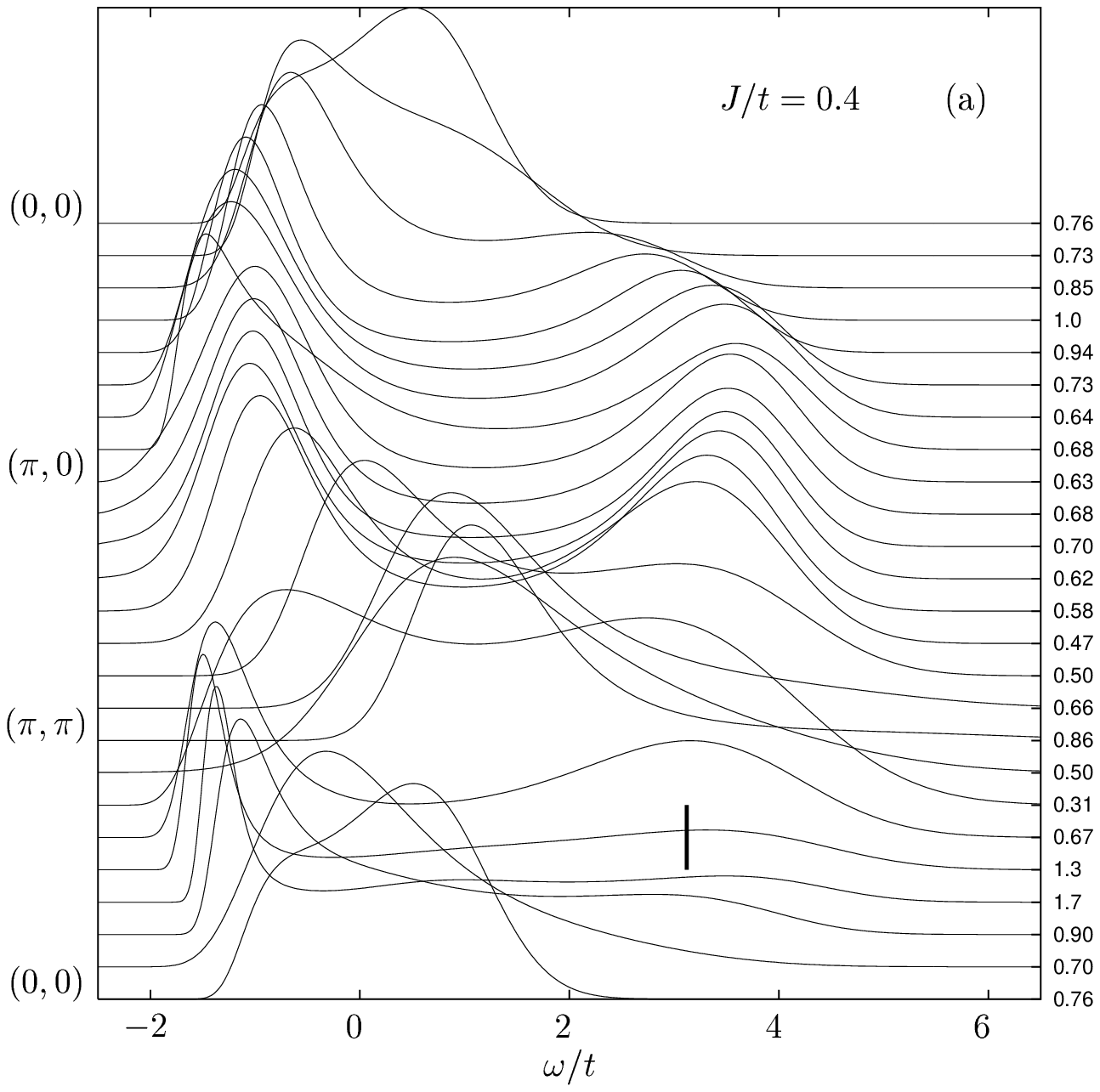,width=6.8cm}                                                

\noindent\epsfig{file=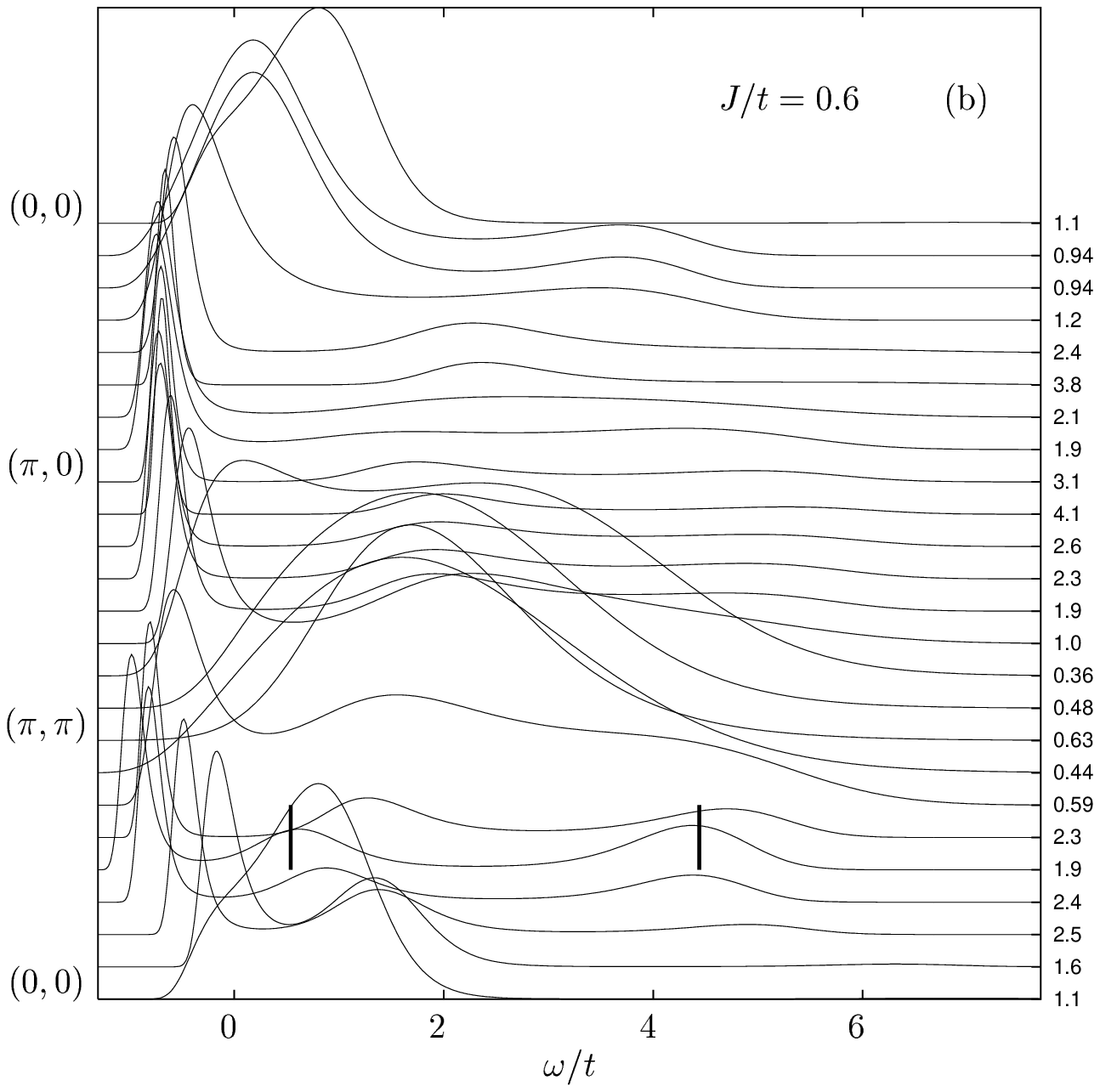,width=6.8cm}                                                         
                                                                                       
\noindent\epsfig{file=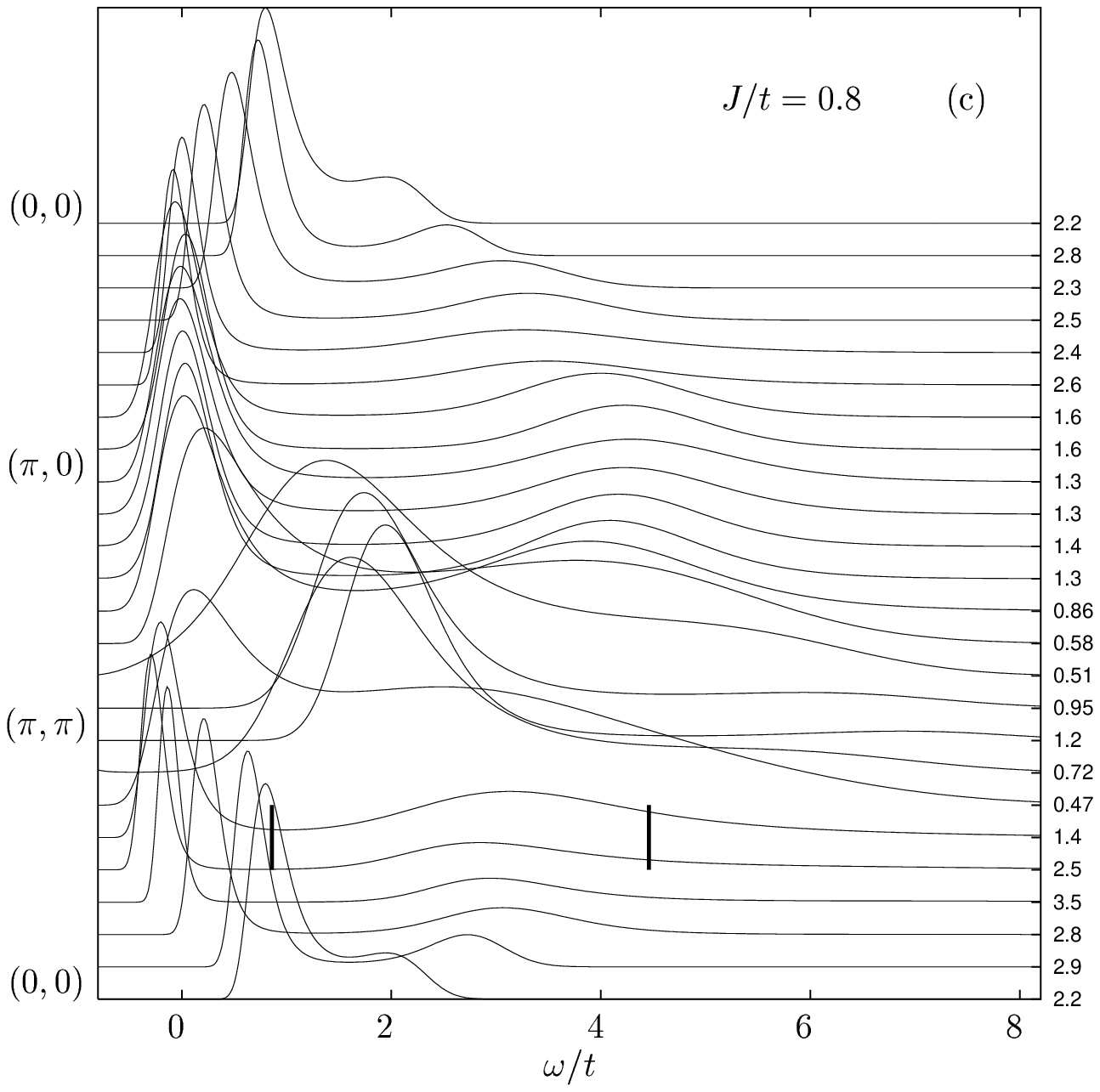,width=6.8cm}                                                

\noindent\epsfig{file=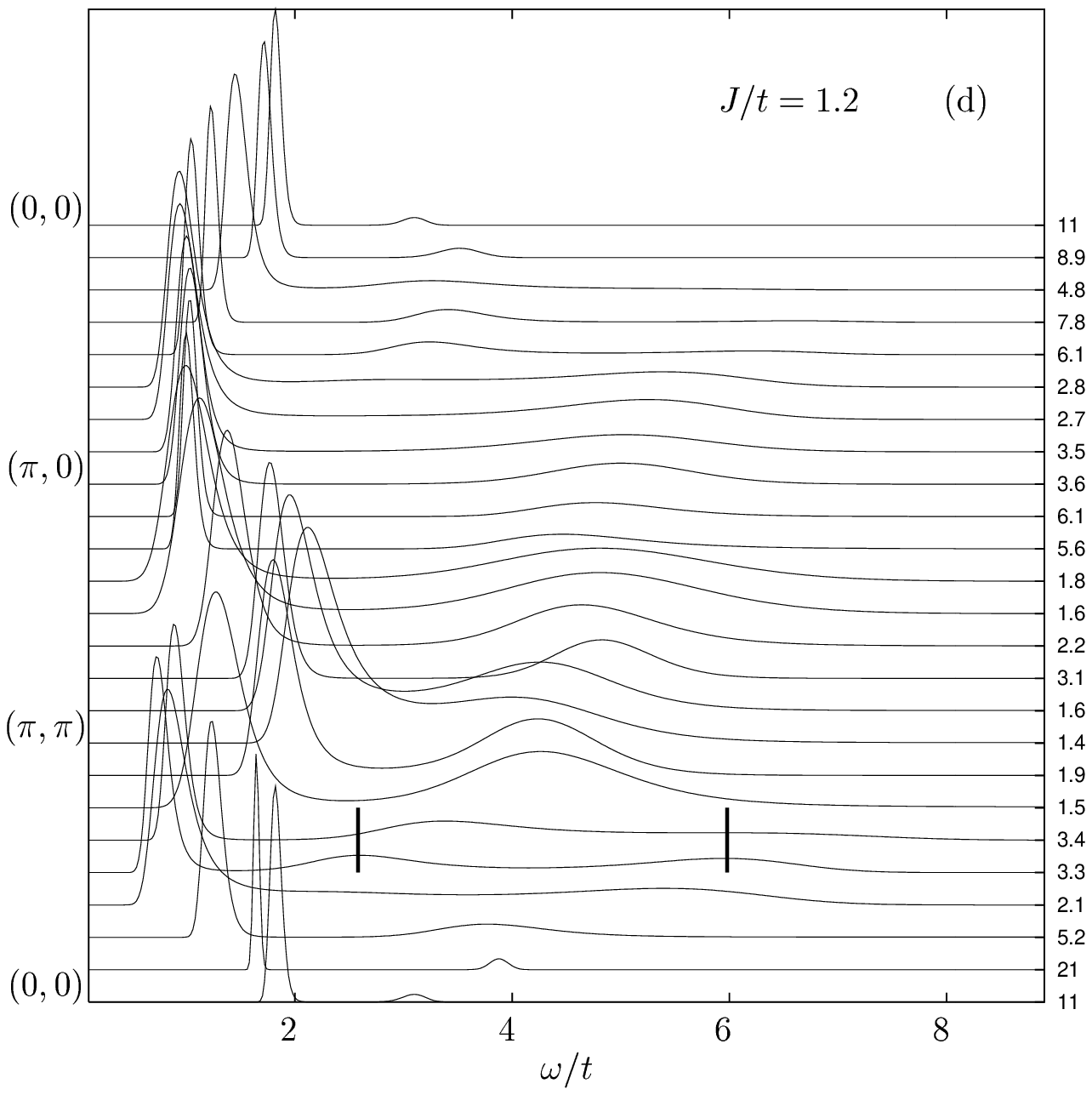,width=6.8cm}                                                         
                                                                                       
\noindent\epsfig{file=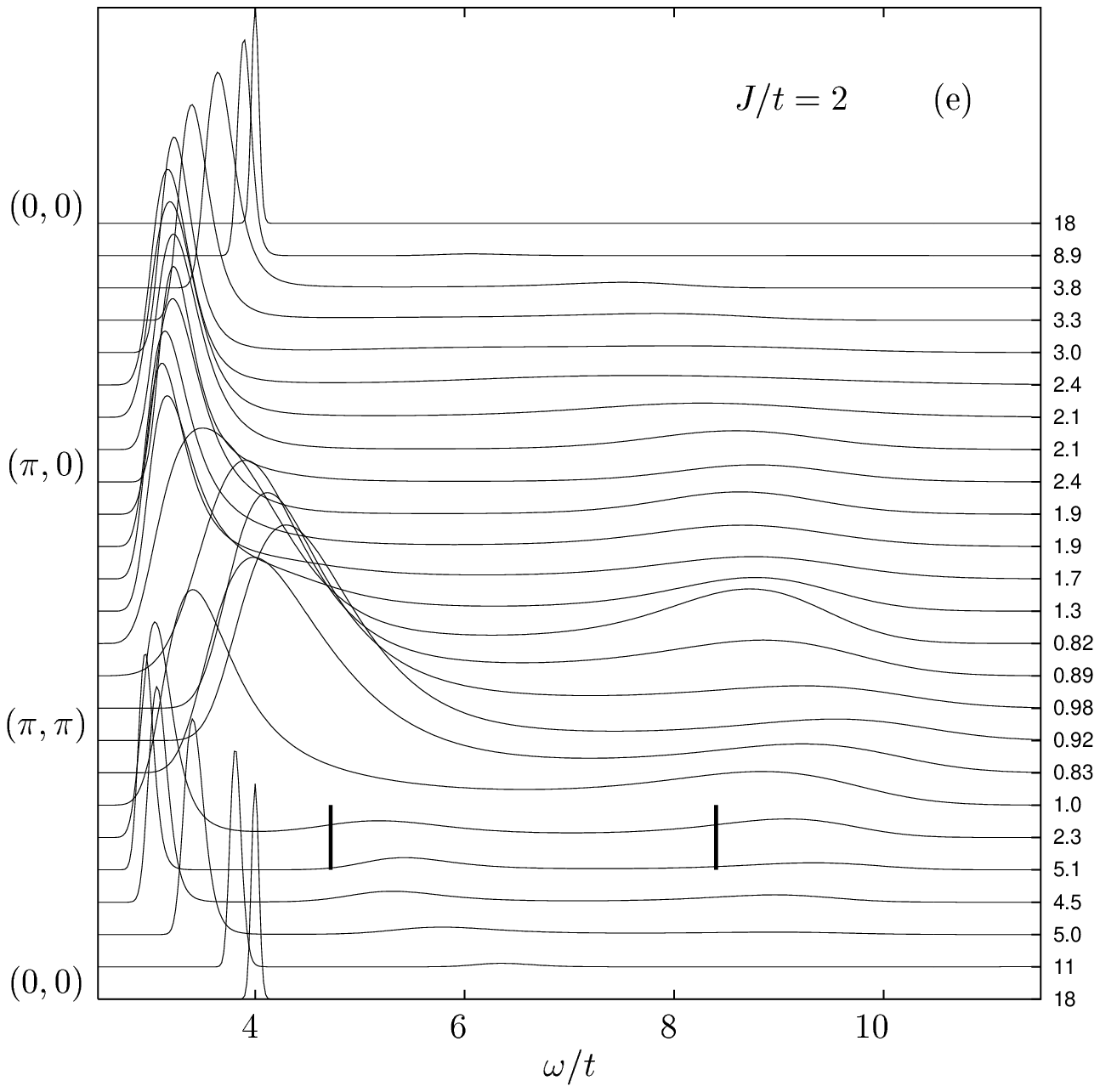,width=6.8cm}                                                

\noindent\epsfig{file=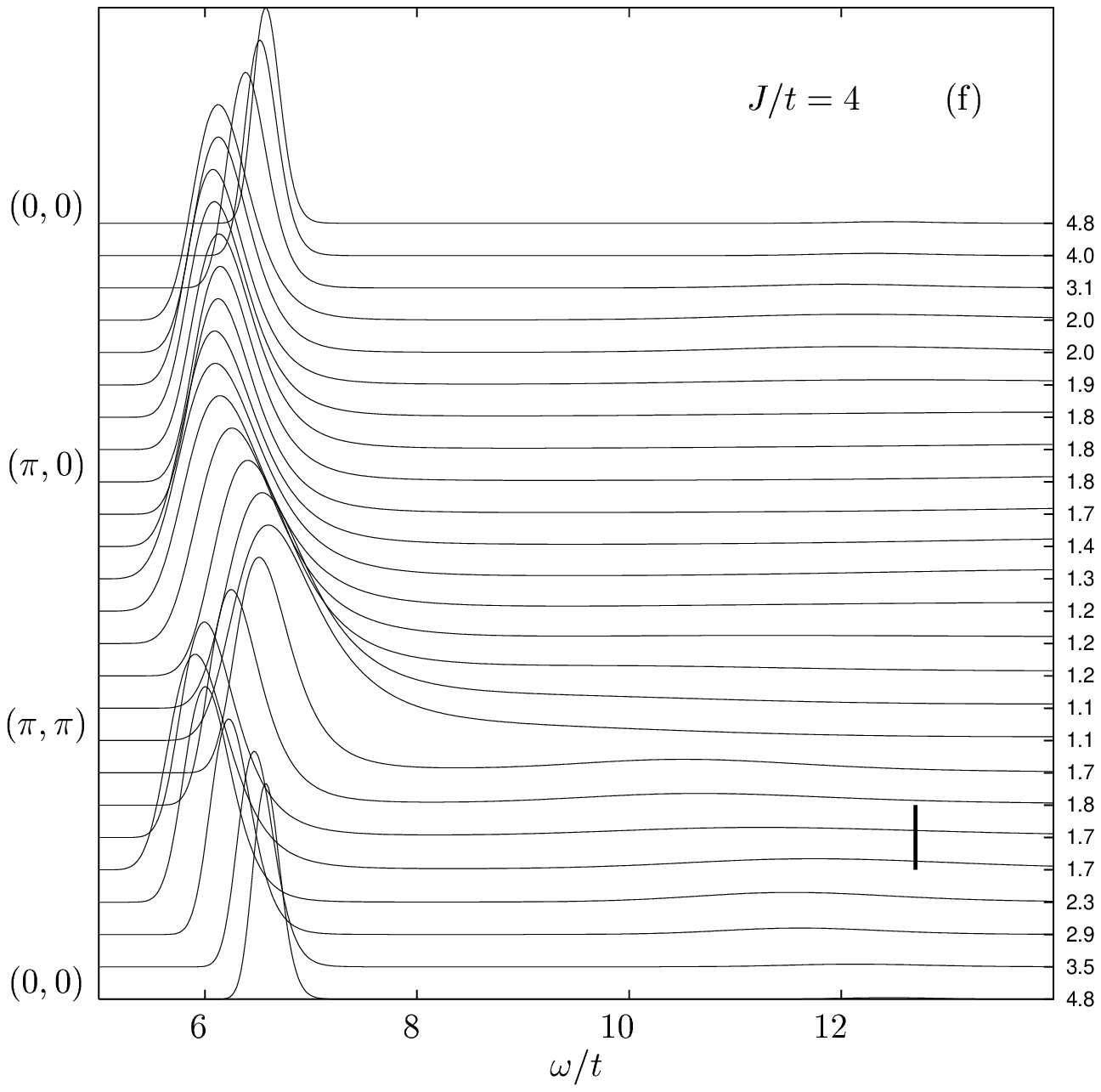,width=6.8cm}                                                         
                                                                                       
\caption{\narrowtext Spectral function for a $16 \times 16$ system        
and a)~$J/t=0.4$, b)~$0.6$,                                                            
c) $0.8$, d) $1.2$, e) $2$ and f) $4$.                                                 
The vertical lines indicate resonances above                                           
the quasiparticle peak at $\vec k=(\pi/2,\pi/2)$ as obtained in                        
Sec.~\ref{sec:sf}, Fig.~\ref{fig:2D2exp}.                                              
\label{fig:Ak2DJ04}}                                                                   
\end{figure}

The results discussed in Sec.\ \ref{sec:low} for the lower edge of the 
spectrum and in Sec.\ \ref{sec:qp} for the 
quasiparticle weight can be recognized in the spectral function
(Fig.~\ref{fig:Ak2DJ04}) obtained by using MaxEnt.
For clarity, the maximum of each
curve is normalized to $1$ in the plots.
The small numbers on the right hand side of the figures correspond to 
the maximal value of $A({\vec k},\omega)$ when the integral
$\int\limits_{-\infty}^{\infty}d\omega A({\vec k},\omega)$ is
properly normalized to $\pi/2$.
The lower edge of the spectrum remains like in the previous section,
but the accuracy of its location in $A({\vec k},\omega)$
is reduced by MaxEnt.
The peaks around $(0,\pi)$ and $(\pi/2,\pi/2)$ are generally very sharp,
in agreement with the fact that a finite quasiparticle weight was found
in Sec.\ \ref{sec:qp}.
A transfer of weight from high to low energies can be observed, 
when $J/t$ is increased, consistent with the increase in 
the quasiparticle weight (Fig.\ \ref{fig:ZJFS} in Sec.\ \ref{sec:qp}). 

When 
compared to the 1D case \cite{brunner99}, 
it is seen that the high energy excitations
in the 2D case are extremely broad.
The total bandwidth remains essentially constant as a
function of $J$ in contrast to
the 1D case, where it scales as $4t+J$.

\begin{figure}[htb]
\noindent\epsfig{file=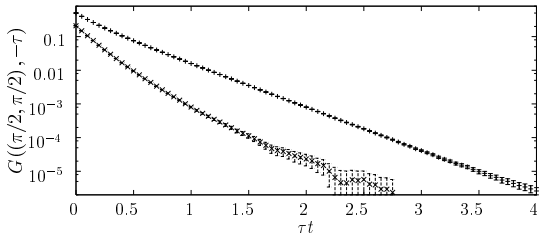,width=8cm}
\caption{\narrowtext Original Green's function and Green's                                         
function with subtraction (lower curve)                                                
for a $16\times 16$ system and $J/t=2$ at ${\vec k}=(\pi/2,\pi/2)$.                    
\label{fig:greensub}}                                                                  
\end{figure}

For values of the coupling in the range $ J/t \le 2$ 
we observe satellite peaks in the region around $\vec k = (\pi/2,\pi/2)$
(Fig.\ \ref{fig:Ak2DJ04}) 
next to the lowest energy peak which is extremely sharp and corresponds 
to a quasiparticle. The $\delta$-peak can not be handled
satisfactorily by MaxEnt.
As can be seen by comparison of Figs.~\ref{fig:2DGF}
and \ref{fig:Ak2DJ04}, MaxEnt
gives some weight at energies lower than the band edge.
This additional weight has to be balanced in some way, such that
this error propagates to the other side of the $\delta$-peak.
Small peaks in the vicinity of the $\delta$-peak can therefore
not be resolved. In order to resolve structures close to the
quasiparticle peak, we subtract the exponential
corresponding to the lowest energy (see Fig.~\ref{fig:greensub}).
The thus modified Green's function can now be used as input of 
MaxEnt.

\begin{figure}                                                                         
\noindent\epsfig{file=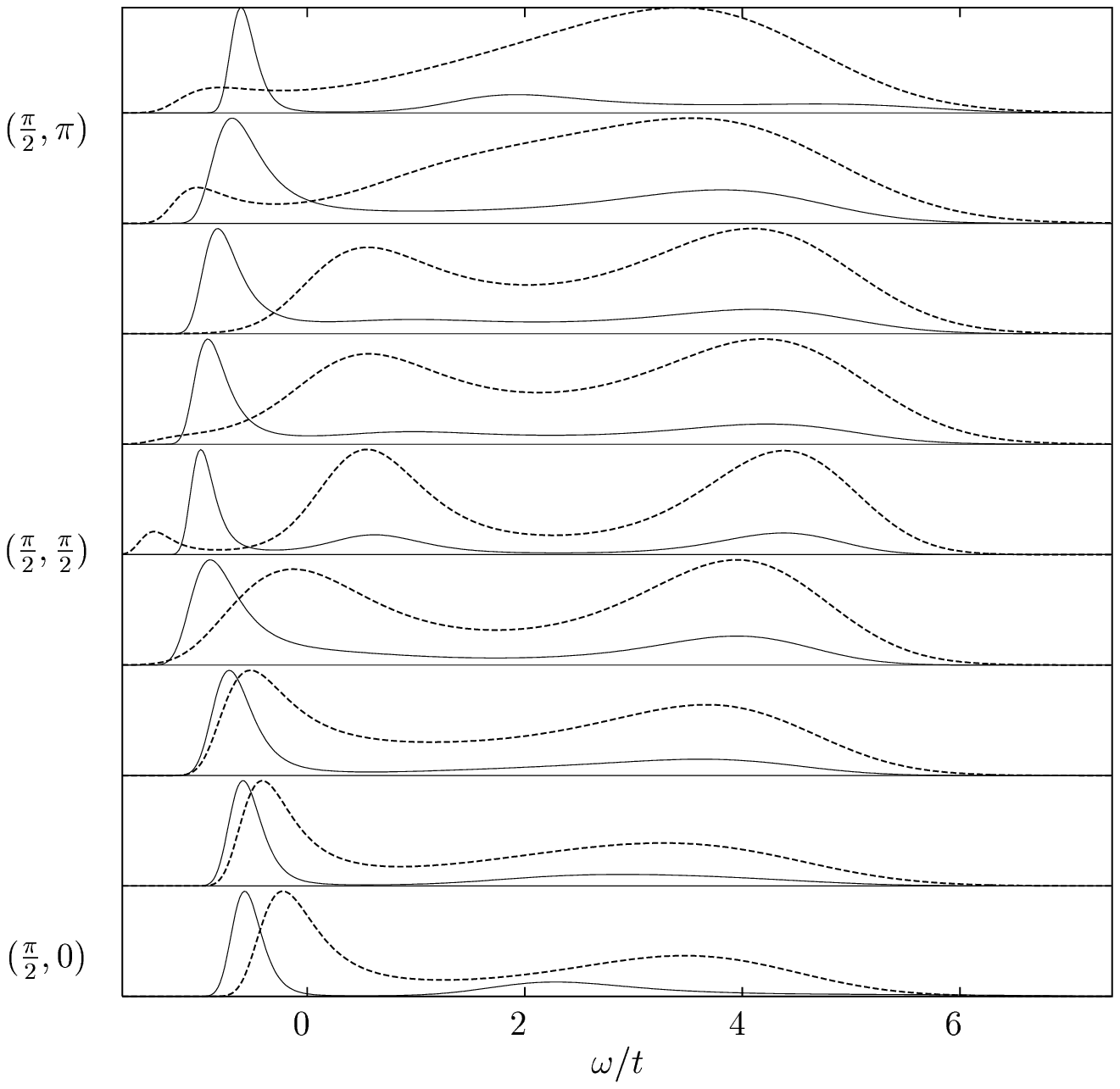,width=7cm}                                                 
\caption{\narrowtext Original spectrum (full line) and after subtraction (dashed line)             
along the line $(\pi/2,k_y)$.                                                          
In the direction toward $(\pi/2,0)$                                                    
the lowest resonance approaches the position                                           
of the quasiparticle peak and merges with it, whereas                                  
toward $(\pi/2,\pi)$ the distance stays approximately constant.                        
In the second case, the main effect is a broadening of the                             
resonances.                                                                            
Shown is a $16 \times 16$ system with $J/t=0.6$.                                       
\label{fig:3exJ06}}                                                                    
\end{figure}                                                                           
                                                                                       
\begin{figure}[hbt]                                                                    
\noindent\epsfig{file=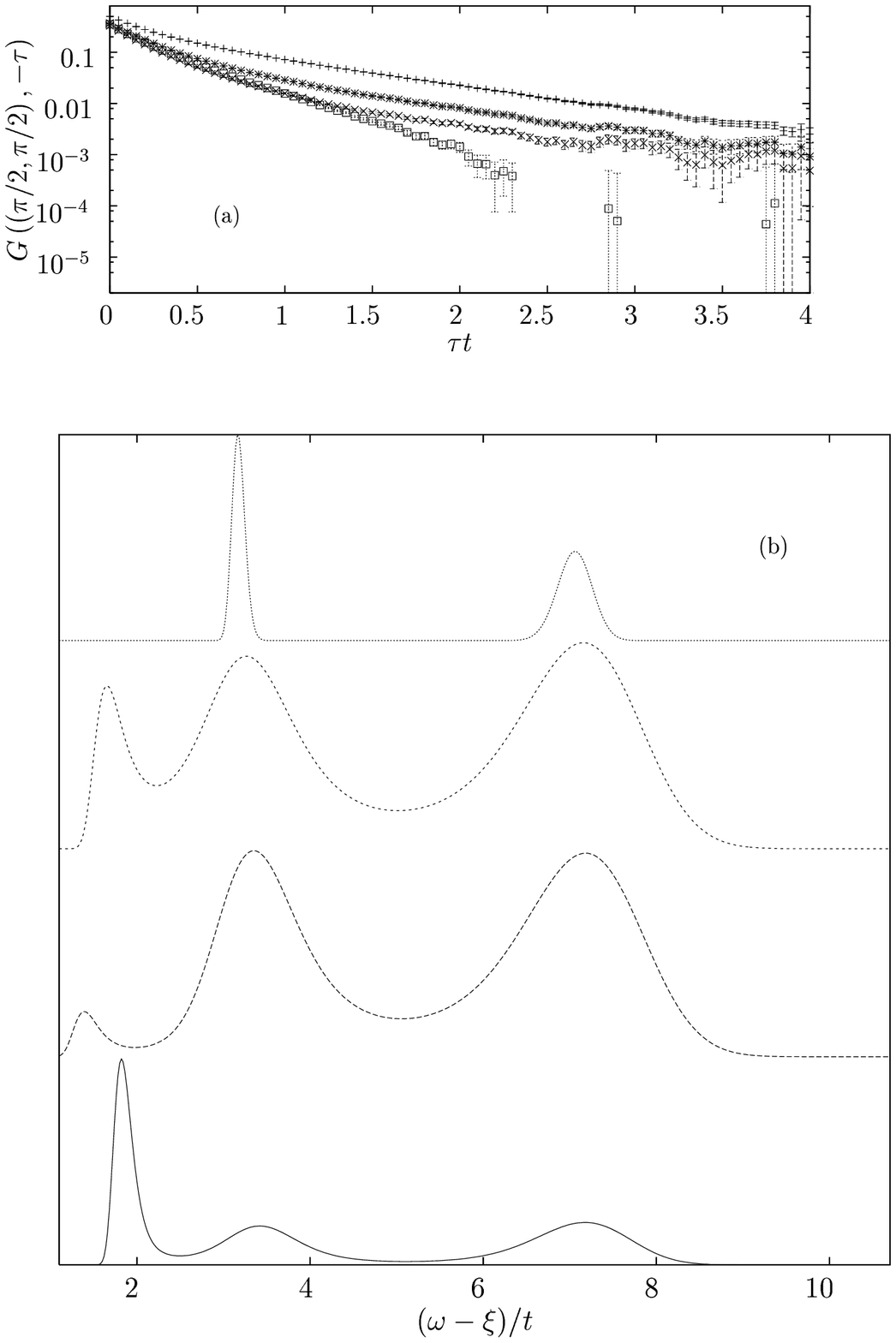,width=8cm}                                                 
\caption{\narrowtext Green's function and resulting spectral function                              
before and after subtraction of the lowest exponential,                                
where the lowest exponential is in the error                                           
bars of section~\ref{sec:low}.                                                         
The results in (b) correspond (from bottom to top)                                     
the original spectral function, the spectral function used in                          
Fig.~\ref{fig:gspi}, the result when subtracting the lowest possible                   
exponential, and to the highest possible exponential. The error bar on                 
the exponential is $0.03 t$, the error on its weight is $0.025$.                       
\label{fig:2DsubJ06test}}                                                              
\end{figure}                                     

Before proceeding to the results, let us remark that, when the  
MaxEnt results obtained with the modified Green's function,
i.e.\ after the subtraction of the lowest exponential,
are viewed closely, 
on occasions, an additional peak appears at
the bottom of the spectrum (this effect can be seen e.g. for
$(\pi/2,\pi/2)$ in Fig.~\ref{fig:3exJ06}).
To exclude, that this peak corresponds to a real physical effect,
we take several modified Green's functions, that are consistent with
the exponential of the lowest peak, within the
statistical error. Therefore, we take the lowest and the highest 
exponential, that are
consistent with the results obtained in Sec.\ \ref{sec:low}, and use
them as input of MaxEnt.
As can be seen in Fig.~\ref{fig:2DsubJ06test}, the new peak that appears
below the low energy peak of the original function,
and hence is artificial,
is only observed in two cases with varying position, whereas the two 
other peaks can be always observed,
no matter which exponential is subtracted (always within the 
statistical errors).
The position of these high energy peaks is not changed
by the different subtractions, only the width is affected. In all cases 
discussed, a small shift of these structures can be observed with respect
to the ones in the spectrum without the subtraction. However, 
the positions assumed by these structures after the subtraction
is not affected 
by the different subtractions within the values allowed by the 
statistical errors. 
We conclude that the initial small shift is due to the inability of
MaxEnt to concentrate the weight of the delta-function
of the quasiparticle peak to a single energy value.

\begin{figure}[tb]                                                                    
\noindent\epsfig{file=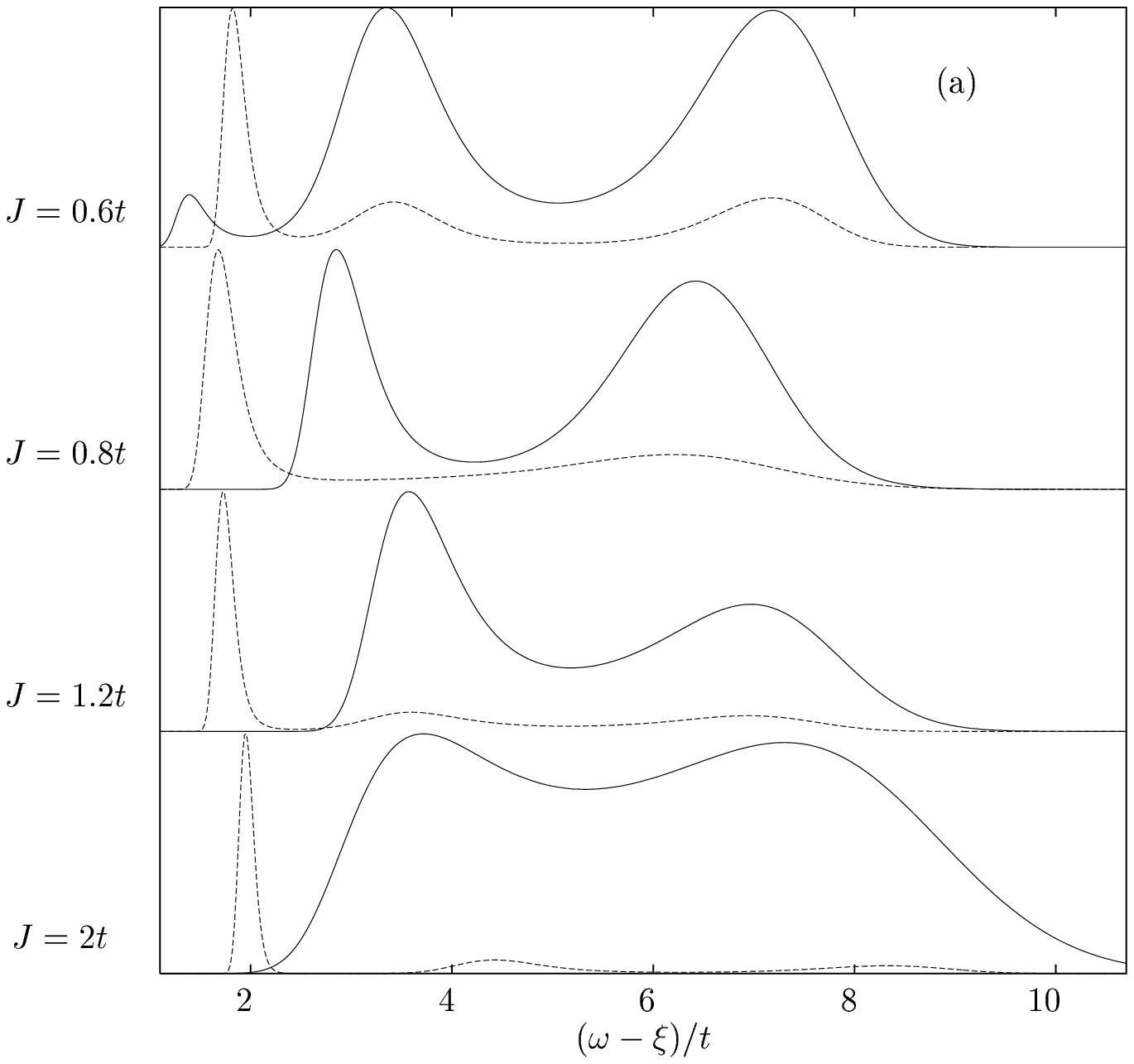,width=8cm}                                                
\epsfig{file=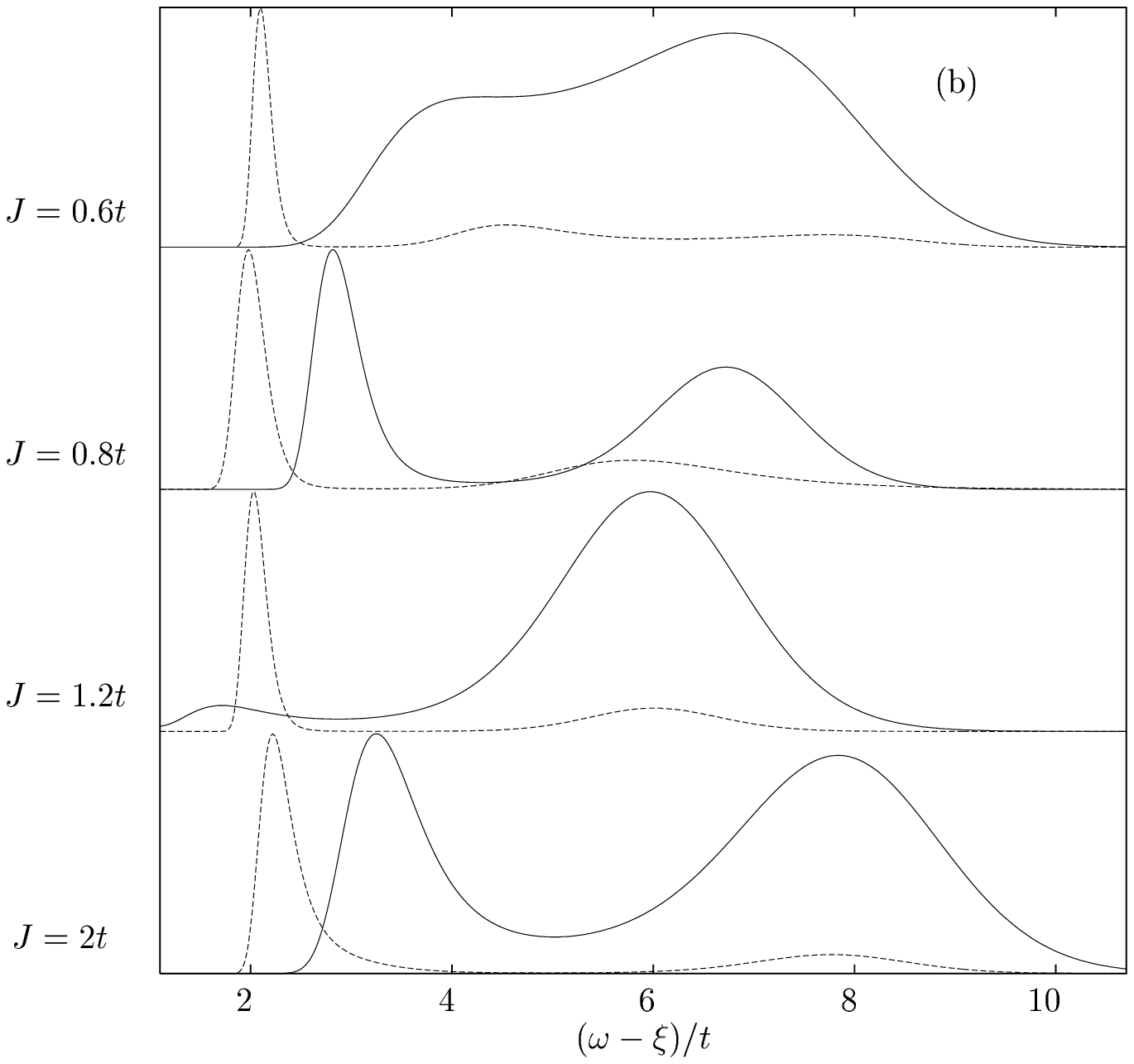,width=8cm}                                                         
\caption{\narrowtext Change of the spectral function when subtracting                              
the first excitation for a) $(\pi/2,\pi/2)$                                            
and b) $(\pi,0)$. The energy is shifted by an arbitrary amount $\xi$,                  
in order to display the spectra for                                                    
different values of $J/t$ in the same energy range.                                    
The dotted line is the original result, the full line gives the modified               
one.                                                                                   
\label{fig:gspi}}                                                                      
\end{figure}         
The  result of the procedure described above
is shown in Fig.~\ref{fig:gspi}.
For $(\pi/2,\pi/2)$ there
are only little changes of the position of the maxima of
the existing peaks
at small $J/t$ compared to the full spectral function (except the low 
energy peak, that disappeared).
We can further observe, that the satellite
peak next to the low energy peak
can now be seen for all values of $J/t\le 2$.
One should notice, that no additional weight has been produced at high 
energies, but the normalization has changed (again the maximal value is 
normalized to 1, not the area of the spectral function).


\begin{figure}[htb]                                                                    
\noindent\epsfig{file=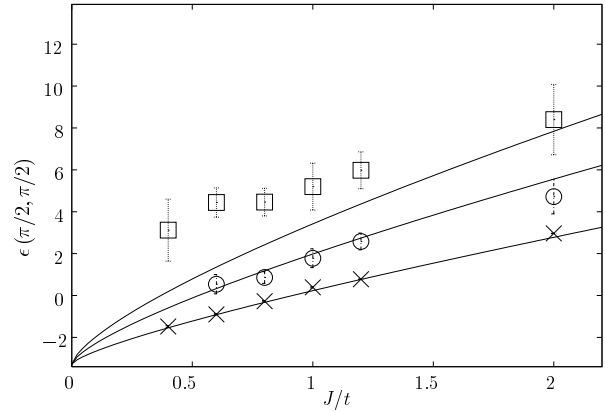,width=8cm}                                                 
\caption{\narrowtext The first three excitations at ${\vec k}=(\pi/2,\pi/2)$.                      
At $J/t=0.4$ only two peaks were resolved.                                             
The lines represent the solutions obtained by solving the linear                       
string potential for the hole in the $t-J_z$ model.                                    
\label{fig:2D2exp}}                                                                    
\end{figure}                        

At $(\pi/2,\pi/2)$ (Fig.~\ref{fig:gspi}(a)) the resolution of the
second-lowest excitation is quite 
clear, when applying the above method,
whereas at $(\pi,0)$ (Fig.~\ref{fig:gspi}(b)) 
the results are 
either not accurate enough, or the corresponding excitation is weaker.
For $J/t=1.2$ the resolution is not good enough to separate 
the two resonances at $(\pi,0)$.
Generally the excitations at higher energies at $(\pi,0)$  are broader
than at $(\pi/2,\pi/2)$, so that the positions of the maxima are not as 
well defined. Similar structures were observed in exact diagonalization
\cite{poilblanc93,dagotto94} and in SCBA \cite{liu92}, and were ascribed 
to string excitations. 

When the string picture is valid, as it is expected
in the $t$-$J_z$ model
the hole is confined by a linear potential, leading to (k-independent)
eigenvalues of the
energy~\cite{bulaevskii68,liu92,elser90} given by
\begin{equation}
E_n/t=-2\sqrt{3}+a_n(J_z/t)^{2/3}
\label{glg:airy},
\end{equation}
where $a_n$ are the eigenvalues of a dimensionless Airy equation 
\cite{kane89}. The first three eigenvalues are given by
$a_n=2.33, 4.08, 5.52$. 
In Fig.~\ref{fig:2D2exp} the results for the first three excitations 
are given for ${\vec k}=(\pi/2,\pi/2)$,
and are compared to the predictions from SCBA.
The error bars on the second and third peak are obtained
as the width of the
MaxEnt peak at half intensity, the error bars of the 
first peak are taken as in Sec.\ \ref{sec:low}.
We find, that for $J/t\le 2$ the lowest peak can be accurately described
by $\epsilon_0 (\pi/2,\pi/2) = -E_H - 3.28 t + 2.33 (J/t)^{2/3}t$, 
where $E_H$ is the Heisenberg energy per site, and the second peak 
by $\epsilon_1 (\pi/2,\pi/2) = -E_H  -3.28 t + 4.08 (J/t)^{2/3}t$.
The value of $3.28 t+E_H$ is the result obtained from
SCBA~\cite{liu92}, whereas the prefactors of $(J/t)^{2/3}$ are
exactly the values of the dimensionless Airy function, implying that
the first two peaks behave (within our error bars) exactly
as it is expected by the string picture. In contrast to this,
a fit from SCBA for the first three excitations
in the $t$-$J$ model for values of $J/t\le 0.4 $ results in
$a_n=2.16, 5.46, 7.81$, also with the exponent $2/3$ \cite{liu92},
leading to a clear disagreement with our data.
The third peak that can be resolved cannot be
explained by the string picture, since
its distance to the lower band edge
is independent of $J$ and has a value of about $4t$. The existence of a 
string excitation is not restricted to
$(\pi/2,\pi/2)$, but it can also be observed
between $(\pi/2,\pi/2)$ and $(\pi/2,3\pi/4)$.
This is demonstrated for the
value $J/t=0.6$ (see Fig.~\ref{fig:3exJ06}).

The results above lead to the conclusion that the lowest excitations
can be well described by the string picture. However, it should be kept
in mind that the string picture originates in the Ising limit for 
$J/t\ll 1$, and that it is based on the continuum limit,
that seems far away from our case with strings
of lengths between two and a maximum of five lattice points,
that correspond to the first two string excitations.
Moreover, the string picture predicts a band without dispersion, that is 
clearly not the case in our simulations. A way to reconcile this
paradoxical situation is given by the very good quantitative agreement
between QMC and series expansions \cite{hamer98} for the dispersion of 
the quasiparticle and its bandwidth for a fairly large range in $J$. 
As shown by the expansion around the Ising
limit, a coherent motion of the hole is made possible after the creation
of strings due to hopping processes, by
appropriate spin-flips, the shortest
string being of length two. The lowest order contribution appears in third
order, where the points $(\pi/2,\pi/2)$ and
$(\pi,0)$ are degenerate. Fourth
and higher order processes remove this degeneracy, giving rise to a band
that agrees qualitatively very well with the one obtained in QMC. 
Therefore on top of the coherent motion determined by $J_{\perp}$,
string like excitations are possible and related to $J_z$ and $t$.
Such a possibility was already proposed by B\'eran, Poilblanc and
Laughlin (BPL) \cite{beran96} on the basis of exact diagonalizations
on small systems and is confirmed
unambiguously by our simulations on large systems.

At $J/t\gtsim 2$ the excitations fall below the values
predicted by the string picture.
In those regions the string picture
is no longer valid, as the relaxation of the disturbed
spin bonds is faster than the motion of the hole.

\section{Conclusions}
\label{sec:conc}
A new QMC algorithm was presented that allows
a rather accurate determination
of the single hole dynamics in a two dimensional Heisenberg $S-1/2$
antiferromagnet. The main advantages of this algorithm are the combination
of the loop algorithm for the update of the spins and the
exact evolution of
the hole for a given spin configuration. Due to the diverging correlation
length at zero temperature, large autocorrelation times should be expected
for algorithms with local updates, a problem that is avoided here by the 
global update of the spins. On the other hand, the exact evolution of the
hole for a given spin background avoids further statistical errors that
would be introduced if the hole is updated stochastically, as in recently 
proposed approaches \cite{prokofev98,brower98}. In fact, the accuracy 
achieved allows for 
a determination of several dynamical quantities on large lattice sizes, 
leading to the possibility of a finite size scaling
of e.g.\ the quasiparticle
weight. 

First (Sec.\ \ref{sec:low}) we discussed the lower edge of the spectrum 
that is obtained
directly from the asymptotics in imaginary time of the Green's function.
This quantity is accessible to different techniques, that are however,
with the exception of GFMC, either restricted to small lattice sizes or 
approximate. The comparison shows that very accurate results are given
by series expansions over a large range of parameters, supporting thus
the interpretation of the relevant physical processes for the coherent
motion of the hole in the frame of a pertubative expansion around the
Ising limit. This picture is further enforced by our study of the 
quasiparticle weight (Sec.\ \ref{sec:qp}) and the spectral function 
(Sec.\ \ref{sec:sf}).
In Sec.\ \ref{sec:qp} it was shown, that indeed
the lower edge of the spectrum
describes the coherent propagation of a hole with finite quasiparticle
weight. This is the case for all the parameter range studied, and due to
the good agreement with SCBA especially for small values of $J$, one can
conclude that this coherent propagation takes place for essentially
$J > 0$. Furthermore, by considering structures next to the lowest peak
in the spectral function (Sec.\ \ref{sec:sf}), it is seen that the lowest 
excitations around
the wave vector $\vec k = (\pi/2,\pi/2)$ are very well described by the
levels of strings usually discussed for the 
$t$-$J_z$ model, giving further support to the pertubative picture, where
the hole creates strings during its motion through the lattice, that
are healed by exchange processes, leading thus to coherence. In fact, 
the strings for the first two levels, that agree
quantitatively with our simulations
correspond to a length of two
and five lattice sites. Strings of length two
are the dominant contributions 
in series expansions for the dispersion of the quasiparticle. 
Moreover, our findings showing the existence of string resonances above
the quasiparticle pole lend support to a picture developed by
BPL \cite{beran96}, where the composite nature of 
the quasiparticle is advanced. In previous exact diagonalization studies, 
the existence of such resonances, that were first observed in $4 \times 4$
lattices \cite{dagotto90} were not clearly identified on larger lattices
\cite{poilblanc93}. We have shown in Sec.\ \ref{sec:sf} that they can be 
quantitatively identified with string excitations.
However, they are visible 
only in a rather narrow region along the line
$k_x \simeq \pi/2$, $\pi/2 \ltsim k_y \ltsim 3\pi/4$, 
such that in small lattices with
up to 26 sites, these features can be very much
affected by boundary effects. 
Following BPL, the quasiparticle can be viewed as a light holon attached
to a spinon by a confining potential, the one that gives
rise to the spectrum
of string excitations.  

A comparison with experiments \cite{wells95,Kim98,Ronning98}
fails 
due to the small gap
between the lowest peak at $\vec k = (\pi/2,\pi/2)$ and the flat
band around $\vec k = (\pi,0)$. It was suggested by several authors
\cite{Kim98,Tohyama99,Martins99} that this shift might be obtained 
introducing hopping terms
to second and third nearest neighbors. Furthermore, it was found in exact 
diagonalizations \cite{Martins99} that such extra terms lead to a 
noticeable reduction of the quasiparticle weight. 
Since exact diagonalizations 
with second and third nearest neighbors in lattices with 18 and 26
sites suffer considerably under finite size effects, a discussion of 
the influence of longer range hopping on the quasiparticle weight must 
be carried out in much larger lattices. Such studies are
presently under way.

This work was supported by Sonderforschungsbereich 382. The numerical
calculations were performed at HLRS Stuttgart. We thank
the above institutions for their support. We are grateful to P.\ Horsch, 
E.\ Manousakis, D.\ Poilblanc, P.\ Prevlo\v{s}ek, and S.\ Sorella
for helpful and instructive discussions, and to CECAM, where part of
these discussions took place, for its hospitality.


\end{multicols}

\end{document}